\documentclass[aip,jcp,a4paper,twoside,reprint,twocolumn,groupedaddress]{revtex4-1}

\usepackage{amsmath}
\usepackage{txfonts}

\usepackage[T1]{fontenc}
\usepackage{textcomp}

\usepackage{bm}
\bmdefine{\bVector}{b}
\bmdefine{\BVector}{B}
\bmdefine{\eVector}{e}
\bmdefine{\EVector}{E}
\bmdefine{\fVector}{f}
\bmdefine{\FVector}{F}
\bmdefine{\gVector}{g}
\bmdefine{\pVector}{p}
\bmdefine{\PVector}{P}
\bmdefine{\qVector}{q}
\bmdefine{\QVector}{Q}
\bmdefine{\rVector}{r}
\bmdefine{\RVector}{R}
\bmdefine{\sVector}{s}
\bmdefine{\uVector}{u}
\bmdefine{\vVector}{v}
\bmdefine{\VVector}{V}
\bmdefine{\muVector}{\mu}
\bmdefine{\OmegaVector}{\Omega}

\usepackage[pdftex]{graphicx}

\usepackage{enumerate}

\begin{document}

\title{Molecular Monte Carlo simulation method of systems connected to three reservoirs}

\author{Yuki Norizoe}
\author{Toshihiro Kawakatsu}
\affiliation{Department of Physics, Tohoku University, 980-8578 Sendai, Japan}

\date{October 20, 2011}

\begin{abstract}
In conventional molecular simulation, metastable structures often survive over considerable computational time, resulting in difficulties in simulating equilibrium states. In order to overcome this difficulty, here we propose a newly devised method, molecular Monte Carlo simulation of systems connected to three reservoirs: chemical potential, pressure, and temperature. Gibbs-Duhem equation thermodynamically limits the number of reservoirs to 2 for single component systems. However, in conventional simulations utilizing 2 or fewer reservoirs, the system tends to be trapped in metastable states. Even if the system is allowed to escape from such metastable states in conventional simulations, the fixed system size and/or the fixed number of particles result in creation of defects in ordered structures. This situation breaks global anisotropy of ordered structures and forces the periodicity of the structure to be commensurate to the system size. Here we connect the such three reservoirs to overcome these difficulties. A method of adjusting the three reservoirs and obtaining thermodynamically stable states is also designed, based on Gibbs-Duhem equation. Unlike the other conventional simulation techniques utilizing no more than 2 reservoirs, our method allows the system itself to simultaneously tune the system size and the number of particles to periodicity and anisotropy of ordered structures. Our method requires fewer efforts for preliminary simulations prior to production runs, compared with the other advanced simulation techniques such as multicanonical method. A free energy measurement method, suitable for the system with the three reservoirs, is also discussed, based on Euler equation of thermodynamics. This measurement method needs fewer computational efforts than other free energy measurement methods do.
\end{abstract}

\maketitle

\section{Introduction}
\label{sec:Introduction}

Metastable structures are found in a variety of physical systems, \textit{e.g.} glasses, amorphous solids of colloids~\cite{Akcora:2009,Norizoe:2005}, stalk intermediate structures in biological membrane fusion process~\cite{Martens:2007,Chernomordik:2008,Norizoe:2010Faraday,DoctoralThesis}, and so on. Metastable states correspond to regions of the local free energy minima in phase space. In conventional simulation of the canonical ensemble, once the system is captured in these regions, the system often remain in these non-equilibrium states for enormously long time. Such metastable states are frequently found in macromolecular and colloidal systems, and make the simulation studies on equilibrium states difficult. Even if the system can escape from the non-equilibrium states to the ordered equilibrium state, the constant number of particles and the constant system size cause defects in the ordered structure. This situation breaks, in a global scale, the anisotropy of the ordered structure and forces the periodicity of the ordered structure to be commensurate to the system size. In order to find defect-free equilibrated ordered structures, we should finely tune the system box size $\left( L_x, L_y, L_z \right)$ as well as the number of particles $N$ so that both the anisotropy and the periodicity of the ordered structure, which are not known \textit{a priori}, are not violated by the periodic boundary conditions. This fine tuning for the ordered structure is in general a tedious task.

Advanced simulation techniques, \textit{e.g.} multicanonical ensemble method~\cite{Berg:1991,Ueda:MolecularSimulationFromClassicalToQuantumMethods} and umbrella sampling~\cite{Torrie:1977,Frenkel:UnderstandingMolecularSimulation2002,Ueda:MolecularSimulationFromClassicalToQuantumMethods}, allow us to almost homogeneously sample the whole phase space at constant $N$ and $\left( L_x, L_y, L_z \right)$, with the help of artificial weights that reduce the occurrence probability of non-equilibrium states. However, microscopic states in equilibrium, sampled by these advanced techniques, are restricted to microstates with the given set of constant $N$ and constant $\left( L_x, L_y, L_z \right)$. Free energy landscapes of the system with different sets of $N$ and $\left( L_x, L_y, L_z \right)$ at the same particle density are not searched. These extensive variables simultaneously need manual fine tuning for the purpose of finding the most stable state in these free energy landscapes. For example, both $N$ and $\left( L_x, L_y, L_z \right)$ of perfect crystals should be integer multiples of the unit structure. Furthermore these advanced methods also require an advanced programming and large amounts of complicated preparation, \textit{e.g.} accurate calculation of free energy~\cite{Lyubartsev:1992} and the precise adjustments of the artificial weights, prior to the production simulation runs. In addition, such unphysical sampling processes with the artificial weights make it difficult to trace physical trajectories in the phase space.

Here we devise molecular Monte Carlo simulation method of systems connected to three reservoirs (hereafter we call it ``three-reservoirs method'')~\cite{2010:NorizoeMuPTLetterArXiv}, chemical potential $\mu$, pressure $P$, and temperature $T$, for seeking the most stable states of the target systems, \textit{i.e.} the equilibrium structures. Due to Gibbs-Duhem equation,
\begin{equation}
\label{eq:Gibbs-Duhem}
	S \, dT - V \, dP + N \, d\mu = 0,
\end{equation}
where $S$ is the entropy and $V$ is the volume, the number of reservoirs is thermodynamically limited to no more than 2 for single component systems. However, we connect the three reservoirs in order to overcome the above difficulties of the other conventional and advanced simulation techniques. In order to perform this, we introduce a method for adjusting these three reservoirs and obtaining thermodynamically stable states.

The total number of particles $N$ and the system box size $\left( L_x, L_y, L_z \right)$ are additional degrees of freedom of the system connected to these three reservoirs. These additional degrees of freedom correspond to additional dimensions of the phase space, which provide shortcuts from the non-equilibrium state to the equilibrium state. In addition, unlike the other simulation techniques utilizing 2 or fewer reservoirs, these degrees of freedom allow the system itself to simultaneously tune $N$ and $\left( L_x, L_y, L_z \right)$, so that the system reaches the true equilibrium ordered structure. Furthermore, our method requires fewer efforts for the preliminary simulation prior to the production simulation runs.

Guggenheim formally introduced Boltzmann factor (statistical weight) of the ensemble with the three reservoirs~\cite{Guggenheim:1939}. Prigogine and Hill also studied the same ensemble later~\cite{Prigogine:1950,Hill:StatisticalMechanicsPrinciplesAndSelectedApplications}. These early works, however, focused on mathematical aspects of the partition function, \textit{i.e.} mathematical formalism of the ensemble, since their goal was to discover a universal and generalized expression for a partition function applicable to any thermodynamically acceptable ensembles~\cite{Sack:1959,Koper:1996}. By contrast, physical aspects of the ensemble were wholly left for the future. In the present work, we study the physical aspects of this ensemble intuitively and thought-experimentally. In addition, we also analytically solve maximization problems of entropy densities, which corroborates our intuitive and thought-experimental study. Finally, we design three-reservoirs method based on these physical aspects of the ensemble and show simulation results on non-trivial globally-anisotropic defect-free ordered structures of colloidal systems.

In spite of Gibbs-Duhem equation, a system connected to the three reservoirs can be realized in experiments. For example, we can imagine a system that obeys the grand canonical ensemble (\textit{i.e.} constant $\mu$, $V$, and $T$), and replace one of its walls with a free piston facing to a reservoir of pressure $P$, \textit{i.e.} the 3rd reservoir. In the present article, we theoretically construct thermodynamics and statistics of the system connected to the three reservoirs.

Based on Euler equation of thermodynamics, we also propose a method for measuring entropy and free energy directly from the simulations of the systems connected to the three reservoirs. This measurement needs fewer computational efforts than the other free energy calculation methods using molecular simulation.

We design the algorithms of the three-reservoirs method based on conventional Monte Carlo (MC) simulation methods of the grand canonical ensemble ($\mu VT$-ensemble) and the isothermal-isobaric ensemble ($NPT$-ensemble). We give a brief description of these conventional molecular MC techniques in appendix~\ref{sec:MolecularMCTechniqueInMuVTAndNPT-ensembles}. Thermodynamics, statistical mechanics, and simulation methods of the system with the three reservoirs are studied in section~\ref{sec:muPTensemble}. Finally, we summarize the present work in section~\ref{sec:Conclusions}.

\section{$\mu PT$-ensemble}
\label{sec:muPTensemble}
Here we discuss the basic formalism of the three-reservoirs method, where the method for adjusting the three reservoirs is also developed. Thermodynamic properties of this system are discussed in section~\ref{subsec:ThermodynamicsInmuPTensemble}. A microscopic formulation of three-reservoirs method based on statistical mechanics will be given in section~\ref{subsec:StatisticalMechanicalPropertiesOfParticlesInmuPTensemble}, where we solve the maximization problem of the statistical entropy per volume. Algorithms for the simulation based on this statistical formulation are constructed in section~\ref{subsec:MCSimulationMethodInmuPTensemble}. Simulation results to demonstrate efficiency and stability of three-reservoirs method are given in section~\ref{subsec:ExaminationOf3-reservoirsMethod}. Finally, a new and simple technique to measure the entropy and the free energy of the system is proposed in section~\ref{subsec:EntropyAndFreeEnergyCalculationInMuPTensemble}.

\subsection{Thermodynamics in $\mu PT$-ensemble}
\label{subsec:ThermodynamicsInmuPTensemble}
As an example of thermodynamic systems, we consider a gas contained in a diathermal box with a free piston. This box is placed in an environment at constant $T$ and constant $P$. These two intensive variables $(T, P)$ determine the other intensive variables of this system, \textit{e.g.} the chemical potential $\mu$, the number density of particles $\rho = N/V$, and the free energy per particle $G/N$. This means that thermodynamic degrees of freedom of this system are equal to 2, which results from Gibbs-Duhem equation, eq.~\eqref{eq:Gibbs-Duhem}. In conventional simulation methods, $N$ is also fixed at some value ($NPT$-ensemble), whereas states and phases of the system are independent of $N$, \text{i.e.} a change of $N$ only scales the extensive variables of the system. In other words, phase diagrams constructed in $PT$-plane are independent of the extensive variables. Instead of fixing this insignificant $N$, we connect this system to a reservoir of $\mu$, whose value is determined by $(T, P)$ through Gibbs-Duhem equation, eq.~\eqref{eq:Gibbs-Duhem}. Since Gibbs-Duhem equation is satisfied, this third reservoir does not affect the equilibrium state of the system. This condition corresponds to a thermodynamically stable point based on Gibbs-Duhem equation, which results in an equation of state that links $T, P$, and $\mu$. At a thermodynamically stable point of this ensemble, the extensive variables of the system, \textit{e.g.} $V$ and $N$, are freely scaled, \textit{i.e.} indeterminate and fluctuating, while the system keeps all the intensive variables fixed. In simulating systems connected to the three reservoirs at the thermodynamically stable points, we can choose simulation runs at small $N$, which are computationally advantageous. Gibbs free energy per particle, which must be minimized in $NPT$-ensemble before adding the 3rd reservoir of constant $\mu$, is unchanged even after this 3rd reservoir is connected to the system.

In the above example, $(T, P)$ is given from the outside of the system; $\mu$ is adjusted according to $(T, P)$ and connected to the system as an additional reservoir. Two other combinations $(T, \mu)$ and $P$, and $(\mu, P)$ and $T$ also work in a similar manner. Therefore, in addition to Gibbs free energy per particle, both grand potential per volume and the thermodynamic potential of $\mu PS$-ensemble per $S$, \textit{i.e.} $(E - \mu N + PV) / S$, are simultaneously minimized in the system connected to the three reservoirs, where $S$ and $E$ are entropy and internal energy respectively.

Gibbs-Duhem equation explains this simultaneous minimization of the 3 free energy densities. When we add the 3rd reservoir to the system and set the system at the thermodynamically stable point, the intensive variable of the 3rd reservoir needs to be adjusted, in advance of the connection of this additional reservoir. Through this adjustment of the 3rd intensive variable, the corresponding free energy density is minimized. The same system at the same thermodynamically stable point is also constructed by the two other combinations of the 3 intensive variables. In other words, $NPT$, $\mu VT$, and $\mu PS$-ensembles simultaneously underlie the ensemble with the three reservoirs. This results in the simultaneous minimization of the 3 free energy densities. In the other ensembles, however, any sets of corresponding 3 external parameters, \textit{e.g.} $(T, V, N)$ in the canonical ensemble ($NVT$-ensemble), can be selected arbitrarily and the adjustment of the external parameters is not demanded. Therefore, the corresponding free energy, \textit{e.g.} Helmholtz free energy in $NVT$-ensemble, is minimized in a conventional ensemble, whereas other free energies are not minimized due to the absence of underlying ensembles.

\begin{figure}[!tb]
	\centering
	\includegraphics[clip]{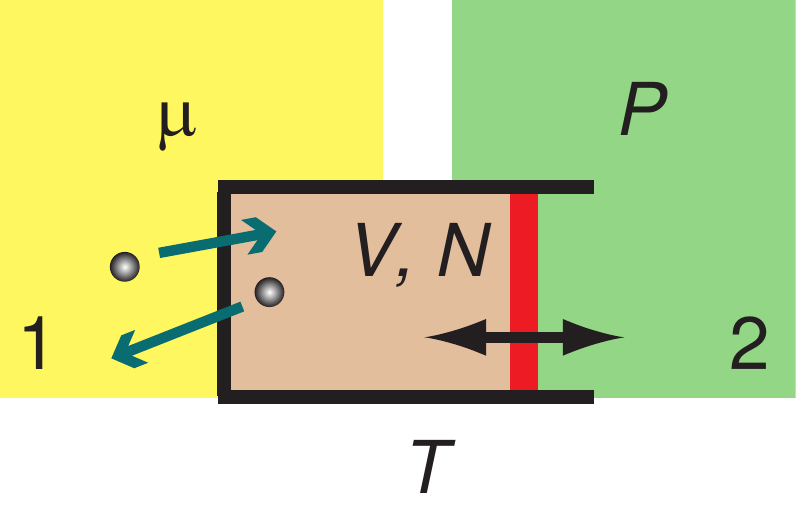}
	\caption{Sketch of a system connected to the three reservoirs. The reservoir 1 and the system are allowed to exchange particles to fix the chemical potential of the system at $\mu$, where $N$ of the system is a dynamic variable. Between the reservoir 2 and the system, a free piston is placed. This piston moves and changes the volume $V$ to fix the pressure of the system at $P$. The system and these reservoirs are connected to a thermostat at $T$, the reservoir 3.}
	\label{fig:SketchMuPTReservoirs}
\end{figure}

A system connected to the three reservoirs is sketched in Fig.~\ref{fig:SketchMuPTReservoirs}. The reservoirs 1 and 2 define the values of $\mu$ and $P$ of the system respectively. The particle density of these 2 reservoirs, denoted by $\rho_{\text{res}}^{(1)}$ and $\rho_{\text{res}}^{(2)}$ respectively, determine the $\mu$ and $P$. The system and the reservoirs 1 and 2 are connected to a thermostat at $T$, the reservoir 3. In appendix~\ref{sec:MolecularMCTechniqueInMuVTAndNPT-ensembles}, we give a derivation of this $\mu$, and its explicit expression is given in eq.~\eqref{eq:ChemicalPotentialOfRealParticles}.

As another simple example, we thermodynamically consider a system composed of a single-component ideal gas. In this example, we assume that the reservoirs are also composed of the same ideal gas. At the thermodynamically stable point, a relation $\rho_{\text{res}}^{(1)} = \rho_{\text{res}}^{(2)}$ holds and the particle number density of the system, $\rho$, also equals these particle number densities of the reservoirs; \textit{i.e.} $\rho_{\text{res}}^{(1)} = \rho_{\text{res}}^{(2)} = \rho$. However, when $\rho_{\text{res}}^{(1)} > \rho_{\text{res}}^{(2)}$, both $N$ and $V$ diverge, since the reservoir 1 continues to increase $N$ of the system aiming for a large $\rho$ value and the reservoir 2 increases $V$ aiming for a small value of $\rho$. When $\rho_{\text{res}}^{(1)} < \rho_{\text{res}}^{(2)}$, both $N$ and $V$ vanish. Therefore, the system reaches equilibrium only at the thermodynamically stable point. In other words, outside the thermodynamically stable point, the system is always in non-equilibrium and both the intensive and the extensive variables are indeterminate. These results also apply to systems composed of interacting particles (\textit{e.g.} non-ideal gases). We utilize these divergence and vanishment of the target system as a criterion for the equilibration, which can be used for the automatic adjustment of the three intensive variables, \textit{e.g.} by a bisection method, to determine the thermodynamically stable point. The system quickly diverges or vanishes outside the vicinity of the thermodynamically stable points~\cite{2010:NorizoeMuPTLetterArXiv}. The speed of the divergence and the vanishment increases with the difference in the intensive parameter sets from the thermodynamically stable point.

On the other hand, when we need a long simulation run in the vicinity of the thermodynamically stable point, indeterminate $N$ could cause a computational problem, since the extensive variables could become extremely large or vanish. However, an appropriate choice of $N$ and $V$ makes the system last for a long time, within which good statistics of simulation results, \textit{e.g.} particle density and lattice constants of crystals, are obtained~\cite{2010:NorizoeMuPTLetterArXiv}. We can determine such simulation results with accuracy enough to obtain the equilibrated structure of the system with the three reservoirs. If we need a far longer simulation run, the ensemble can be switched to one of the conventional methods, \textit{e.g.} $NPT$-ensemble or $NVT$-ensemble. As these switched ensembles are free from the problem of the indeterminate $N$, they allow us to perform a longer simulation run of the equilibrated structure obtained via three-reservoirs method.

In the present article, we tentatively call the ensemble of the systems connected to the three reservoirs $\mu PT$-ensemble, since this is obtained as an equilibrium condition between the three intensive variables and as a combination of $\mu VT$, $NPT$, and $\mu PS$-ensembles.

\subsection{Statistical mechanical properties of particles in $\mu PT$-ensemble}
\label{subsec:StatisticalMechanicalPropertiesOfParticlesInmuPTensemble}
Here the statistical mechanical properties of particles in $\mu PT$-ensemble are discussed. According to Gibbs-Duhem equation, the thermodynamic potential of $\mu PT$-ensemble is identically equivalent to zero in the thermodynamic limit, whereas systems obeying this ensemble have certain degrees of freedom in the phase space. This means that Boltzmann factor fluctuates in statistical mechanics and that the partition function of this ensemble, \textit{i.e.} summation of statistical weights over the whole phase space, is defined. This Boltzmann factor is calculated in the present section as a natural extension of those for $\mu VT$ and $NPT$-ensembles, and determines detailed balance conditions necessary for designing three-reservoirs simulation method.

For determining this Boltzmann factor, we solve a maximization problem of the statistical entropy per volume in $\mu PT$-ensemble. The statistical entropy is defined as~\cite{Sack:1959,Reichl:ModernCourseInStatisticalPhysics},
\begin{equation}
\label{eq:DefinitionStatisticalEntropy}
	-k_B \sum_j p_j \log p_j,
\end{equation}
where the suffix, $j$, represents microstates of the system, $p_j$ denotes corresponding occurrence probability, and $k_B$ is Boltzmann constant. Because the extensive variables of $\mu PT$-ensemble is indeterminate, we choose the entropy density to be maximized. With the use of eq.~\eqref{eq:DefinitionStatisticalEntropy}, the statistical entropy per volume is given by,
\begin{equation}
\label{eq:EntropyPerVolume}
	-k_B \sum_j \frac{ p_j \log p_j }{ V_j }.
\end{equation}
The equilibrium probability distribution, $\{ p_1, p_2, \dots, p_j, \dotsc \}$, which maximizes eq.~\eqref{eq:EntropyPerVolume} under the constraints in $\mu PT$-ensemble, is determined by:
\begin{gather}
	\label{eq:EntropyMaxConstraintsNormalization}
	\sum_j p_j = 1,  \\
	\label{eq:EntropyMaxConstraintsEnergyAndParticleDensity}
	\sum_j \left( p_j \frac{ E_j }{ V_j } \right) = \left\langle \frac{ E }{ V } \right\rangle,  \qquad
	\sum_j \left( p_j \frac{ N_j }{ V_j } \right) = \langle \rho \rangle,
\end{gather}
where $\langle \dotsm \rangle$ is the thermal average specified according to the reservoirs. The constraint of eq.~\eqref{eq:EntropyMaxConstraintsNormalization} represents the normalization condition. The two constraints given by eq.~\eqref{eq:EntropyMaxConstraintsEnergyAndParticleDensity} instead of three constraints are due to the thermodynamic degrees of freedom, which equal 2. With the use of Lagrange multipliers, $\tilde{\alpha}$, $\tilde{\beta}$, and $\tilde{\gamma}$, this maximization problem is reduced to~\cite{Reichl:ModernCourseInStatisticalPhysics},
\begin{align}
\label{eq:EntropyMaximizationProblem}
	\max_{\{ p_j \}} &\left[ -k_B \sum_j \frac{ p_j \log p_j }{ V_j } + \tilde{\alpha} \left( \sum_k p_k - 1 \right) \right.  \notag \\
	&\left. + \tilde{\beta} \left( \sum_k p_k \frac{ E_k }{ V_k } - \left\langle \frac{ E }{ V } \right\rangle \right) + \tilde{\gamma} \left( \sum_k p_k \rho_k - \langle \rho \rangle \right) \right].
\end{align}
This reduced problem, eq.~\eqref{eq:EntropyMaximizationProblem}, is solved by a partial derivative with respect to $p_j$. The solution is:
\begin{equation}
\label{eq:EntropyMaxSolution}
	p_j = \exp \left[ \alpha V_j - \beta E_j + \gamma N_j - 1 \right],
\end{equation}
where $\alpha = \tilde{\alpha} / k_B$, $\beta = -\tilde{\beta} / k_B$, and $\gamma = \tilde{\gamma} / k_B$. Equations~\eqref{eq:EntropyPerVolume} - \eqref{eq:EntropyMaxConstraintsEnergyAndParticleDensity}, and \eqref{eq:EntropyMaxSolution} give the statistical entropy per volume determined as thermal average as:
\begin{equation}
\label{eq:EntropyPerVolumeAsThermalAverage}
	\frac{ S }{ V } = -k_B \left\{ \alpha - \beta \left\langle \frac{ E }{ V } \right\rangle + \gamma \left\langle \frac{ N }{ V } \right\rangle - \left\langle \frac{ 1 }{ V } \right\rangle \right\}.
\end{equation}
Equating eq.~\eqref{eq:EntropyPerVolumeAsThermalAverage} with the thermodynamic relation for the thermodynamic potential of $\mu PT$-ensemble, denoted by $\Phi$,
\begin{equation}
\label{eq:ThermodynamicPotentialMuPT}
	\Phi = E- TS - \mu N + PV,
\end{equation}
we obtain:
\begin{gather}
\label{eq:DeterminedLagrangeMultipliers}
	\alpha = - \frac{ P }{ k_B T },  \qquad  \beta = \frac{ 1 }{ k_B T },  \qquad  \gamma = \frac{ \mu }{ k_B T },  \\
\label{eq:StatisticalThermodynamicPotentialMuPT}
	\Phi = -k_B T \log e.
\end{gather}
In the thermodynamic limit, the right-hand side of eq.~\eqref{eq:StatisticalThermodynamicPotentialMuPT} is essentially zero compared with the other extensive variables in eq.~\eqref{eq:ThermodynamicPotentialMuPT}. This result coincides with the Euler equation in thermodynamics. Furthermore, for finite $V$, the last term of eq.~\eqref{eq:EntropyPerVolumeAsThermalAverage} is monotonically increasing with decreasing $V$. This indicates that due to the principle of maximizing entropy a finite size system as is used in the computer simulation has a tendency to shrink even at the thermodynamically stable point. Finally, from eq.~\eqref{eq:EntropyMaxSolution}, the probability distributions of the microstates are obtained as,
\begin{equation}
\label{eq:StatisticalProbabilityDensityMuPT}
	p_j = \frac{ 1 }{ e } \exp \left[ - \frac{ P }{ k_B T } V_j - \frac{ 1 }{ k_B T } E_j + \frac{ \mu }{ k_B T } N_j \right].
\end{equation}
We confirmed that these results are also obtained by maximizing the statistical entropy per particle or the statistical entropy per internal energy instead of the statistical entropy per volume as has been done in eq.~\eqref{eq:EntropyMaximizationProblem}.

The Boltzmann factor of $\mu PT$-ensemble is,
\begin{equation}
\label{eq:BoltzmannFactorMuPT}
	\exp \left[ - \frac{ 1 }{ k_B T } \left\{ PV - \mu N + U \left( \rVector_1 , \dots , \rVector_N \right) + E_{\text{kinetic}} \right\} \right],
\end{equation}
where $E_{\text{kinetic}}$ denotes the total kinetic energy of the system, $U$ the potential energy of the system, and $\rVector_i$ is the spatial coordinates of the particle $i$. As we have seen above, this is a result of a natural extension of $\mu VT$ and $NPT$-ensembles. This Boltzmann factor determines the statistical properties of systems equilibrated with the three reservoirs and is equivalent to the Boltzmann factor of $NPT$-ensemble at fixed $N$, and is equivalent to the Boltzmann factor of $\mu VT$-ensemble at fixed $V$.

\subsection{Thermodynamic potential of $\mu PT$-ensemble}
\label{subsec:ThermodynamicPotentialOfMuPT-ensemble}
Guggenheim formally derived the Boltzmann factors (statistical weight) of various ensembles~\cite{Guggenheim:1939}. Assuming that the ensemble averages of extensive variables, e.g. $V$ and $N$, were determined in $\mu PT$-ensemble, Guggenheim also introduced statistical weight of this ensemble, based on analogy between other conventional ensembles~\cite{Guggenheim:1939,Koper:1996}. However, in the calculation of the maximization of entropy density discussed in section~\ref{subsec:StatisticalMechanicalPropertiesOfParticlesInmuPTensemble}, Guggenheim's assumption corresponds to keeping the averages $\langle E \rangle$ and $\langle N \rangle$ fixed, instead of the constraints eq.~\eqref{eq:EntropyMaxConstraintsEnergyAndParticleDensity}. Guggenheim's assumption contradicts the indetermination of the extensive variables, as was pointed out by Prigogine and further discussed by Sack~\cite{Prigogine:1950,Sack:1959}. Prigogine showed that the summation of the resulting Boltzmann factor over the phase space diverges and therefore concluded that the resulting partition function of $\mu PT$-ensemble does not have any physical meanings~\cite{Prigogine:1950,Sack:1959}. The thermodynamic potential of $\mu PT$-ensemble, $\Phi$, determined from such partition function could be indefinite while it should identically equal zero in the thermodynamic limit. The true thermodynamic potential that dominates this ensemble, similar to the Helmholtz free energy in $NVT$-ensemble, remains unknown since Guggenheim's article. In the following, we will answer Prigogine's criticism and give an explicit expression of the thermodynamic potential for $\mu PT$-ensemble.

In conventional ensembles, the statistical weight takes non-zero values only in the vicinity of the averages $\langle N \rangle$ or $\langle V \rangle$ in the phase space. Outside this vicinity, the statistical weight quickly decreases to zero. This suppresses the divergence of the partition functions, \textit{i.e.} the summation of the statistical weights, in the case of conventional ensembles. On the other hand, in $\mu PT$-ensemble, there is no such limitation because of the indeterminate extensive variables. The statistical weight of $\mu PT$-ensemble at each microstate keeps non-zero values at any $N$ or $V$. This results in the divergence of the partition function of $\mu PT$-ensemble. However, ratios of the statistical weights between any pair of microstates are still defined. This feature guarantees the physical validity of $\mu PT$-ensemble. In this case, the trajectory of the system in the phase space is similar to a free random walk in infinitely large space without boundaries.

Moreover, in the present study, equations~\eqref{eq:StatisticalThermodynamicPotentialMuPT} and \eqref{eq:StatisticalProbabilityDensityMuPT} calculated with the constraints eqs.~\eqref{eq:EntropyMaxConstraintsNormalization} and \eqref{eq:EntropyMaxConstraintsEnergyAndParticleDensity} indicate that the summation of the statistical weights equals $e$ for $\mu PT$-ensemble. Therefore, the corresponding thermodynamic potential $\Phi$, eq.~\eqref{eq:StatisticalThermodynamicPotentialMuPT}, is negligibly small compared with the other extensive variables in eq.~\eqref{eq:ThermodynamicPotentialMuPT} and vanishes in the thermodynamic limit. Furthermore, by the thermodynamic consideration given in section~\ref{subsec:ThermodynamicsInmuPTensemble}, we have shown that $\mu PT$-ensemble is obtained by combining 3 underlying ensembles each with 2 reservoirs, \textit{i.e.} $NPT$, $\mu VT$, and $\mu PS$-ensembles. This thermodynamic consideration means that free energy densities of these underlying ensembles, \textit{i.e.} Gibbs free energy per particle, grand potential per volume, and $(E - \mu N + PV) / S$, are simultaneously minimized in $\mu PT$-ensemble, rather than $\Phi$. This corresponds to the minimization of Helmholtz free energy in $NVT$-ensemble.

\subsection{MC simulation method in $\mu PT$-ensemble}
\label{subsec:MCSimulationMethodInmuPTensemble}
The present simulation method is constructed based on conventional simulation methods in the grand canonical ensemble ($\mu VT$-ensemble) and $NPT$-ensemble. The simulation algorithms of the particle insertion and deletion in $\mu VT$-ensemble (see appendix~\ref{subsec:MCSimulationMethodInMuVTensemble}) and the system size change in $NPT$-ensemble (see appendix~\ref{subsec:MCSimulationMethodInNPTensemble}) are directly utilized in our method. This compatibility between the present method and the conventional methods demonstrates that the algorithms of the present method satisfy detailed balance condition.

One simulation step of the present method is composed of the following 4 trial steps:
\begin{enumerate}[i)]
	\item with probability $p_G / 2$, trial particle insertion into the system,
	\item with probability $p_G / 2$, trial particle deletion from the system,
	\item with probability $p_V$, trial system size change,
	\item with probability $1 - p_G - p_V$, trial displacement by Metropolis algorithm \textit{i.e.} perturbation of one particle,
\end{enumerate}
is chosen, where $p_G$ and $p_V$ are constants fixed in an interval $0 \le p_G, p_V \le 1$.

With the use of the simulation algorithms of trial particle insertion and deletion in $\mu VT$-ensemble, the insertion and deletion of the present method, steps i) and ii), are performed. During this particle exchange between the system and the reservoir 1, the system size $(L_x, L_y, L_z)$ is fixed. This particle exchange in the present method satisfies the detailed balance condition because it is guaranteed in $\mu VT$-ensemble.

The trial system size change, step iii), is performed with use of the simulation algorithms in $NPT$-ensemble, during which $N$ is fixed. This system size change in the present method also satisfies the detailed balance condition. Unlike the conventional MC simulations in $NPT$-ensemble based on McDonald's method~\cite{Frenkel:UnderstandingMolecularSimulation2002}, $L_x, L_y$, and $L_z$ are independently changed in the present method.

The trial move of one particle, step iv), is performed by Metropolis algorithm in $NVT$-ensemble, which also satisfies the detailed balance condition.

Therefore, the present MC simulation method for $\mu PT$-ensemble fulfills the principle of detailed balance. See also section~\ref{subsubsec:DetailedBalanceConditionAndErgodicityOf3-reservoirsMethod}.

Our algorithm indicates that a short-time average of an intensive physical quantity in $\mu PT$-ensemble is approximated by ensemble averages of $\mu VT$ and $NPT$-ensembles at corresponding $N$ and $V$, which is discussed in appendix~\ref{sec:EnsembleAverageAtEachNAndVInmuPTensemble}.

Our simulation is performed in a rectangular system box with independently changing system size $(L_x, L_y, L_z)$. Since any crystals fit into rectangular boxes with periodic boundary conditions, we do not have to introduce Parrinello-Rahman method~\cite{Parrinello:1980,Parrinello:1981,Najafabadi:1983,Frenkel:UnderstandingMolecularSimulation2002}, which allows the change in the shape of the simulation box.

Our simulation algorithm is similar to Gibbs ensemble technique, which is utilized for simulation of phase equilibria in $NVT$-ensemble~\cite{Panagiotopoulos:1987,Panagiotopoulos:1988,Ueda:MolecularSimulationFromClassicalToQuantumMethods,Frenkel:UnderstandingMolecularSimulation2002}, where the phases coexisting in the same system box in $NVT$-ensemble exchange both volume and particles. A system connected to the three reservoirs at a thermodynamically stable point corresponds to this Gibbs ensemble, when one of the coexisting phases in Gibbs ensemble is assumed to be infinitely large that plays the role of the two reservoirs $\mu$ and $P$.

\subsubsection{Detailed balance condition and ergodicity of three-reservoirs method.~~}
\label{subsubsec:DetailedBalanceConditionAndErgodicityOf3-reservoirsMethod}
In this section, the detailed balance condition and the ergodicity of three-reservoirs method are discussed. Steps i) and ii) change $N$ according to the detailed balance condition as in the same way that the $\mu VT$-ensemble does. Step iii) changes $(L_x, L_y, L_z)$ according to the detailed balance condition as in the $NPT$-ensemble. The particle coordinates are updated in step iv), by the standard Metropolis algorithm of $NVT$-ensemble. As a result, three-reservoirs method satisfies the principle of detailed balance and ergodicity, based on the conventional ensembles which fulfill ergodicity. This indicates that $N$, $\left( L_x, L_y, L_z \right)$, and the particle coordinates are simultaneously updated in the phase space, so that the system realizes the equilibrium state. This also means that statistics of $\mu PT$-ensemble contradicts none of the three underlying ensembles with 2 reservoirs, \textit{i.e.} $\mu VT$, $NPT$, and $\mu PS$-ensembles.

\subsection{Examination of three-reservoirs method}
\label{subsec:ExaminationOf3-reservoirsMethod}
In this section, we demonstrate the efficiency and stability of the three-reservoirs method using several examples.

The first example is a model polymer-grafted colloidal system, which was observed to show various exotic metastable phases each of which has a long life time~\cite{Norizoe:2005}.

Colloidal particles are made from metals, polymers, \textit{etc.} and are often modeled with hard spheres~\cite{Eral:2009}. On the other hand, owing to van der Waals attraction acting on particle surfaces, colloids are aggregating and make precipitates after a long time. For the purpose of stabilizing colloidal dispersions against the precipitation, linear polymer chains are often grafted onto the surfaces of the colloids. These are called polymer-grafted colloids. Depending on the physical and chemical properties of the grafted polymers, interaction between polymer-grafted colloids changes significantly~\cite{Akcora:2009}. Polymer-grafted colloids have several industrial applications due to such useful characteristics, for example filler particles immersed in a polymer matrix, and the particles in electro and magnet rheological fluids.

In our previous work~\cite{Norizoe:2005}, we studied the phase behavior of colloidal particles onto which diblock copolymers are grafted. Pair interaction potential between these polymer-grafted colloids was numerically determined via self-consistent field calculation~\cite{Norizoe:2005} as a function of the distance between centers of the particles, $r$. This potential has been approximated by spherically symmetrical repulsive square-step potential with a rigid core of diameter $\sigma_1$ and a square-step repulsive potential of diameter $\sigma_2$ and height $\epsilon_0$ as:
\begin{alignat*}{2}
 & \phi (r) = \infty              & \qquad & r < \sigma_1,  \\
 & \phi (r) = \epsilon_0 \; (>0) &        & \sigma_1 < r < \sigma_2, \\
 & \phi (r) = 0                   &        & \sigma_2 < r.
\end{alignat*}
where the step potential is originated from the grafted polymer brushes. This interaction potential $\phi (r)$ is purely repulsive. Simulating particles interacting via $\phi (r)$ in $NVT$-ensemble, we have studied phase behavior of these colloidal systems. These MC simulation results~\cite{Norizoe:2005} show that, at low temperature, high pressure, and $\sigma_2 / \sigma_1 \approx 2$, our particles self-assemble into string-like assembly. The positions and the mean-square displacement of the particles show that the string-like assembly is observed in disordered solid phases. Actually, such a string-like assembly has been observed recently in experiments~\cite{Osterman:2007}. In addition to such string-like assembly, various structures, \textit{e.g.} dimers and lamellae~\cite{Malescio:2003,Glaser:2007}, and also glass transition~\cite{Fomin:2008} are observed in the same model system. It was also shown that the particles interacting via continuous repulsive potential similar to the above $\phi (r)$ show these string-like and other various assemblies~\cite{Camp:2003}. In these recent studies using $NVT$-ensemble at finite $T$ in both 2 and 3-dimensions, it was shown that this string-like assembly with a local alignment in the same direction but with a global isotropy is a metastable structure. Although a variety of ground states of the same model system at zero temperature have been discovered via genetic algorithms~\cite{Pauschenwein:2008}, equilibrium states at finite temperature have not been understood yet.
\begin{figure*}[!tb]
	\centering
	\includegraphics[clip]{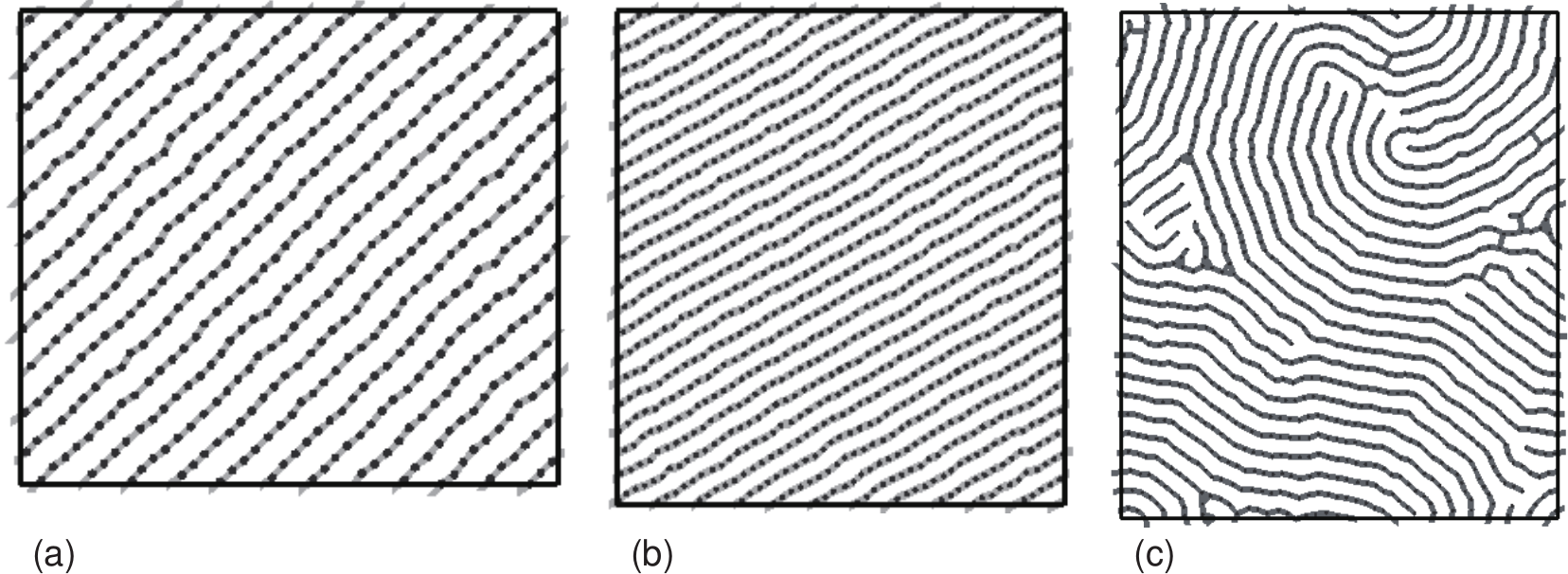}
	\caption{Snapshots of the system simulated in $\mu PT$-ensemble.
	(a), $k_B T / \epsilon_0 = 0.12 $, $P \sigma_1^{2} / \epsilon_0 = 1.0$ and $\mu' = 27.63$. 
	(b), $k_B T / \epsilon_0 = 0.12 $, $P \sigma_1^{2} / \epsilon_0 = 1.1$ and $\mu' = 29.93$. 
	Black dots represent the centers of the particles and grey lines denote networks of overlaps between the particles. A snapshot of the system simulated in $NVT$-ensemble at $k_B T / \epsilon_0 = 0.12$, $\rho \sigma_1^2 = 0.451$, and $N = 1200$ is also shown in (c)~\cite{Norizoe:2005}.}
	\label{fig:Frog2DmuPTS20Snapshots}
\end{figure*}

We simulate the equilibrated states of our model system at finite $T$ via three-reservoirs method. In the present simulation work, $\sigma_1$ and $\epsilon _0$ are taken as the unit length and the unit energy, respectively. Simulation is performed on 2 dimensional systems. We define dimensionless chemical potential as~\cite{DoctoralThesis},
\begin{equation}
\label{eq:DefinitionDimensionlessChemicalPotential}
	\mu' := \mu / k_B T - \log \left( \left. \varLambda^2 \right/ \sigma_1^2 \right).
\end{equation}
The thermal de Broglie wave length $\varLambda$, defined in eq.~\eqref{eq:ThermalDeBroglieWaveLength} in appendix~\ref{sec:MolecularMCTechniqueInMuVTAndNPT-ensembles}, is removed from $\mu$ in eq.~\eqref{eq:DefinitionDimensionlessChemicalPotential}, since simulation results are independent of $\varLambda$, which is discussed in appendix~\ref{sec:MolecularMCTechniqueInMuVTAndNPT-ensembles}. In the present article, $\sigma_2 / \sigma_1 = 2.0$ is fixed.

In the initial state, $N_0$ particles are arranged on a homogeneous triangular lattice in a square system box, $L_x / L_y = 1.00$, with the periodic boundary condition. For the trial move of the particles, \textit{i.e.} Metropolis algorithm, a particle is picked at random and given a uniform random and isotropic trial displacement within a square whose sides have length $0.4 \sigma_1$. $\varDelta L$ is fixed at $0.01 \sigma_1$. The probability $p_G = 0.1$ and $p_V = 1 / N_0$. With this $p_V$, the computational time for the simulation is about twice as long as the simulation in $NVT$-ensemble. 1 Monte Carlo step (MCS) is defined as $N_0$ simulation steps. The Mersenne Twister algorithm~\cite{MersenneTwister1,MersenneTwister2,MersenneTwister3} is adopted as a uniform random number generator for our simulation.

\subsubsection{String-like assembly.~~}
\label{subsubsec:String-likeAssembly}
In our previous $NVT$-ensemble simulation with the potential step width, $\sigma_2 / \sigma_1 = 2.0$, we found~\cite{Norizoe:2005} that the string length diverges at $k_B T / \epsilon_0 = 0.12$ and $\rho \sigma_1^2 \approx 0.451$. At this low $k_B T / \epsilon_0$, using three-reservoirs method, we simulate the system. In the present simulation, the density $\rho \sigma_1^2$ is initially set at 0.451 for $N_0 = 1254$ system.

\begin{figure}[!tb]
	\centering
	\includegraphics[clip]{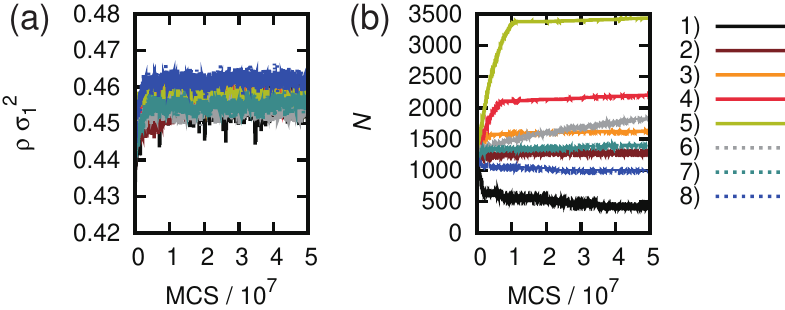}
	\caption{Time evolution of $\rho = N/V$, (a), and time evolution of $N$, (b), simulated at $k_B T / \epsilon_0 = 0.12$ and various $( P \sigma_1^{2} / \epsilon_0$, $\mu' )$. Lines 1) to 5) are simulation results at $P \sigma_1^{2} / \epsilon_0 = 1.0$.
		1): 
			$\mu' = 27.63$.
		2): 
			$\mu' = 29.93$.
		3): 
			$\mu' = 31.54$.
		4): 
			$\mu' = 32.23$.
		5): 
			$\mu' = 34.53$.
		Lines 6) to 8) are simulation results at $\mu' = 29.93$.
		6): $P \sigma_1^{2} / \epsilon_0 = 0.95$. 
		7): $P \sigma_1^{2} / \epsilon_0 = 0.98$. 
		8): $P \sigma_1^{2} / \epsilon_0 = 1.1$. 
	}
	\label{fig:Frog2DmuPTS20T012ForArticle-RhoAndN}
\end{figure}

First, simulating the system at various values of $\mu'$, we search for the thermodynamically stable point at fixed $P \sigma_1^{2} / \epsilon_0 = 1.0$. The given $\mu'$ is, 1): $\mu' = 27.63$, 2): $\mu' = 29.93$, 3): $\mu' = 31.54$, 4): $\mu' = 32.23$, and 5): $\mu' = 34.53$. As an example, a snapshot of the system at 1) is presented in Fig.~\ref{fig:Frog2DmuPTS20Snapshots}(a). Despite different $\mu'$, all the systems of 1) to 5) show similar well-aligned globally-anisotropic defect-free string-like assembly, though the system in $NVT$-ensemble at $\rho \sigma_1^2 = 0.451$ and $N = 1200$ presents the globally-isotropic string-like assembly as is shown in Fig.~\ref{fig:Frog2DmuPTS20Snapshots}(c)~\cite{Norizoe:2005}. Only small and short-lived defects caused by the thermal fluctuation can be generated in the systems simulated with $\mu PT$-ensemble, whereas many long-lived defects are observed in $NVT$ simulation. Time evolutions of $\rho \sigma_1^2$ and $N$ at 1) to 5) are given in Fig.~\ref{fig:Frog2DmuPTS20T012ForArticle-RhoAndN}. All the data for the time evolution of $\rho \sigma_1^2$ shown in Fig.~\ref{fig:Frog2DmuPTS20T012ForArticle-RhoAndN}(a) are fluctuating in the vicinity of $\rho \sigma_1^2 \approx 0.451$, regardless of $\mu'$. It would be worth noting that the simulations started from different initial conditions, \textit{e.g.} different initial particle density $\rho \sigma_1^2$ and different system aspect ratio of the simulation box $L_x / L_y$, also reach the same results as long as the intensive variables of the reservoirs, $\mu'$, $P \sigma_1^{2} / \epsilon_0$, and $k_B T / \epsilon_0$ are the same. Furthermore, outside this range of $\mu'$ from 1) to 5), the system diverges ($N \to \infty$) or vanishes ($N \to 0$) just after the simulation starts. These results illustrate that this region of $\mu'$ is located in the vicinity of the thermodynamically stable point at this pressure, $P \sigma_1^{2} / \epsilon_0 = 1.0$.

On the other hand, time evolution of $N$, plotted in Fig.~\ref{fig:Frog2DmuPTS20T012ForArticle-RhoAndN}(b), indicates that the total system size depends on $\mu'$ in a systematic manner. At small MCS, line 1) indicates the tendency of the vanishment and lines 2) to 5) the tendency of the divergence. This means that, in a short computational time, the thermodynamically stable point is expected to lie between 1): $\mu' = 27.63$ and 2): $\mu' = 29.93$, which corresponds to a relative error of some percent. The abrupt time evolution of $N$ stops at less than $10^7$ MCS, whereas the system slowly continues diverging or vanishing at larger MCS. The MCS needed for the divergence or the vanishment becomes larger when we choose parameter sets close to the exact values at the thermodynamically stable point. This is utilized as a criterion for measuring the convergence of the thermodynamic intensive variables in $\mu PT$-ensemble. Although we stop the simulation with this relative error of some percent, the accuracy of the thermodynamically stable point could be raised, \textit{e.g.} by the bisection method improving the accuracy of $\mu'$.

Lines 6) to 8) in Fig.~\ref{fig:Frog2DmuPTS20T012ForArticle-RhoAndN} show the results of a similar series of simulations for a fixed value of $\mu' = 29.93$ (the same value as that for line 2)) and changing the value of $P \sigma_1^{2} / \epsilon_0$. This $\mu'$ is consistent with the thermodynamically stable point at $P \sigma_1^{2} / \epsilon_0 = 1.0$ obtained above within the relative error of some percent. The given $P \sigma_1^{2} / \epsilon_0$ of lines 6) to 8) ranges $P \sigma_1^{2} / \epsilon_0 = 0.95$ to 1.1. A snapshot of the system at 8) is presented in Fig.~\ref{fig:Frog2DmuPTS20Snapshots}(b). Despite different intensive parameter sets, all the systems of 6) to 8) show the globally-anisotropic defect-free string-like assembly similar to Fig.~\ref{fig:Frog2DmuPTS20Snapshots}(a). Long-lived defects are absent in these systems. All the lines 6) to 8) in Fig.~\ref{fig:Frog2DmuPTS20T012ForArticle-RhoAndN}(a) are fluctuating around lines 1) to 5), \textit{i.e.} the vicinity of $\rho \sigma_1^2 \approx 0.451$. For the values of $P \sigma_1^{2} / \epsilon_0$ outside this range, $N$ quickly diverges or vanishes. From these data, we recognize that the present thermodynamically stable point lies between 7): $P \sigma_1^{2} / \epsilon_0 = 0.98$ and 8): $P \sigma_1^{2} / \epsilon_0 = 1.1$. This thermodynamically stable point is equal to the one we obtained above within a relative error of some percent, as is expected.

Once the thermodynamically stable point is determined accurately in $\mu PT$-ensemble, we can exchange the ensemble to a conventional one, \textit{e.g.} $NPT$-ensemble or $NVT$-ensemble, and can perform a longer simulation run, which is free from the divergence or the vanishment of $N$. As is expected, we can perform longer simulation runs even with the $\mu PT$-ensemble if the thermodynamically stable point is determined with higher accuracy.

Next, we check the stability of our simulation method. Starting from the instantaneous microstate shown in Fig.~\ref{fig:Frog2DmuPTS20Snapshots}(a), we resume simulation runs after disconnecting the reservoirs 1 or 2, which corresponds to simulations with $NPT$ or $\mu VT$-ensemble.

\begin{figure}[!tb]
	\centering
	\includegraphics[clip]{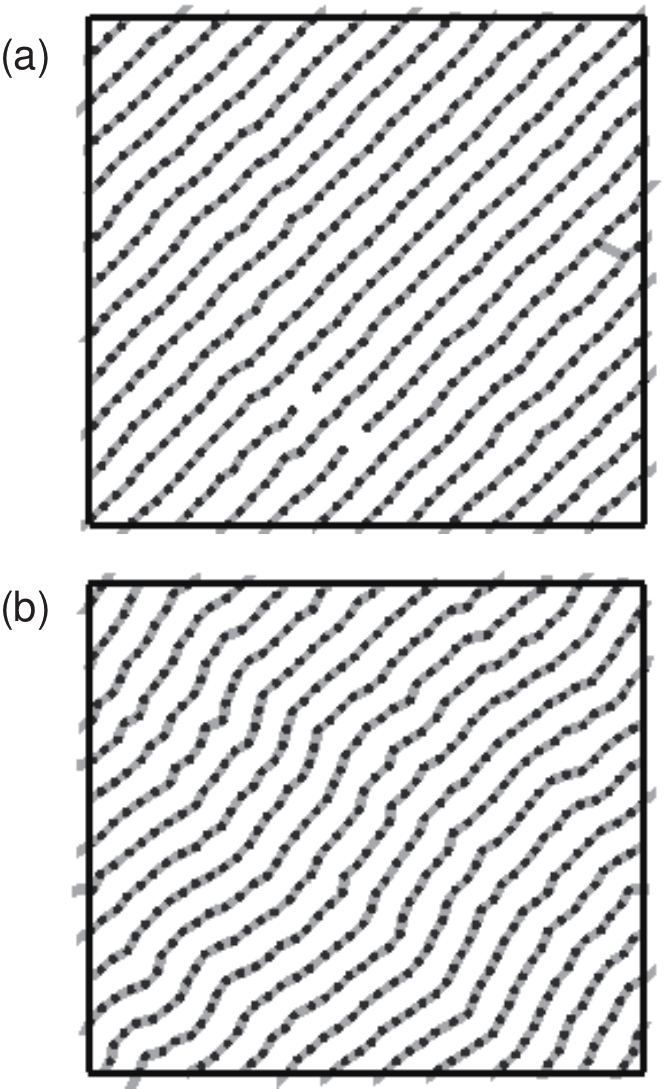}
	\caption{Snapshots of the model polymer-grafted colloidal system at $k_B T / \epsilon_0 = 0.12$. Black dots represent the centers of the particles and grey lines denote networks of overlaps between the particles.
	From the instantaneous state of Fig.~\ref{fig:Frog2DmuPTS20Snapshots}(a), simulations are resumed after disconnecting the reservoir 1 or 2, i.e. simulations are resumed with $NPT$ or $\mu VT$-ensemble. (a): simulation result at $1.1 \times 10^7$ MCS, after resuming the simulation with $NPT$-ensemble. (b): simulation result at $1.6 \times 10^7$ MCS, after resuming the simulation with $\mu VT$-ensemble.
	}
	\label{fig:Frog2DmuPTS20SnapshotsInVariousEnsembles}
\end{figure}

\begin{figure}[!tb]
	\centering
	\includegraphics[clip]{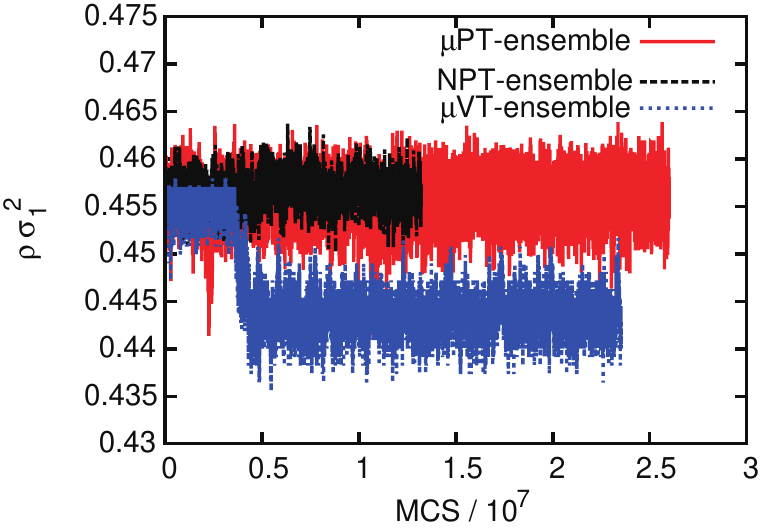}
	\caption{Time evolution of $\rho \sigma_1^2$ in simulations with $NPT$ and $\mu VT$-ensembles that are resumed from the instantaneous state shown in Fig.~\ref{fig:Frog2DmuPTS20Snapshots}(a). In these simulation runs, the number of the particles is $N \approx 500$ with minor fluctuations. For reference, the result of the simulation with $\mu PT$-ensemble is also shown.}
	\label{fig:Frog2DmuPTS20RD001e14P10T012InNPTmuVT-rho}
\end{figure}

Snapshots of the system and the time evolution of $\rho \sigma_1^2$ in these resumed simulation runs are given in Figs.~\ref{fig:Frog2DmuPTS20SnapshotsInVariousEnsembles} and \ref{fig:Frog2DmuPTS20RD001e14P10T012InNPTmuVT-rho} respectively. Both of these simulation runs preserve the same string-like assembly. In $NPT$-ensemble case, a value of $\rho \sigma_1^2$ that is close to that for three-reservoirs method is also obtained. Therefore, the equilibrium state obtained via three-reservoirs method does not change even after the ensemble is switched. These results justify our method for determining the thermodynamically stable point with $\mu PT$-ensemble.

Here, one can recognize occurrence of a few defects and undulation of the string-like assemblies after switching the ensembles. This should be attributed to the difference in the nature of the thermal fluctuations of a finite system between different ensembles. For example, when a particle is removed from a perfect triangular crystal in $\mu PT$-ensemble and simultaneously the ensemble is exchanged to $NPT$-ensemble, long-lived defects are created in the crystal. Figure~\ref{fig:Frog2DmuPTS20SnapshotsInVariousEnsembles}(a) shows this finite size effect. In the thermodynamic limit where the system size becomes infinitely large, such a difference should vanish. We also recognize a slight drop in $\rho \sigma_1^2$ in $\mu VT$-ensemble by approximately 2\%. This is also related to the finite size effect of the $\mu VT$-ensemble, which will be further discussed in section~\ref{subsubsec:AlderTransitionOfTheOuterCores}. The undulation of the string-like assembly shown in Fig.~\ref{fig:Frog2DmuPTS20SnapshotsInVariousEnsembles}(b) corresponds to zigzag instability typically observed in convection rolls in a fluid slab when its natural periodicity is suddenly changed~\cite{Cross:1993}. This zigzag instability is consistent with the slight drop in $\rho \sigma_1^2$, \textit{i.e.} a slight increase in the system size, in $\mu VT$-ensemble shown in Fig.~\ref{fig:Frog2DmuPTS20RD001e14P10T012InNPTmuVT-rho}.

These simulation results show that the physical properties of the system, in conventional ensembles, sensitively depend on $N$ and the system box size while they do not strongly depend on $N$ in $\mu PT$-ensemble.

\subsubsection{Alder transition of the outer cores.~~}
\label{subsubsec:AlderTransitionOfTheOuterCores}
At extremely low temperature, the repulsive square-step of $\phi (r)$ becomes far higher than the thermal energy, $k_B T$. Due to this extremely high potential energy barrier, the phase behavior of the system is almost identical to the behavior of hard particle systems with diameter $\sigma_2$ if the system volume $V$ exceeds the close-packed volume (area) of the outer cores of the particles, denoted by $V_0$ ($= N {\sigma_2}^2 \sqrt3 /2$ in 2 dimensions)~\cite{Norizoe:2005}. Therefore, crystalization of the hard particles at high density, called Alder transition~\cite{Alder1962}, occurs in our system at $V / V_0 \approx 1.3$ and low temperature~\cite{Norizoe:2005}. Here we simulate this triangular crystal of the outer cores of our colloids with the diamter $\sigma_2$ at low temperature $k_B T / \epsilon_0 = 0.1$, where we fixed the parameters $(P \sigma_1^{2} / \epsilon_0 = 0.45, \mu' = 18.42 )$ that are determined via the iterative refinement of $P$ and $\mu$. The stability of this thermodynamically stable point is discussed later in the present section.
\begin{figure}[!tb]
	\centering
	\includegraphics[clip]{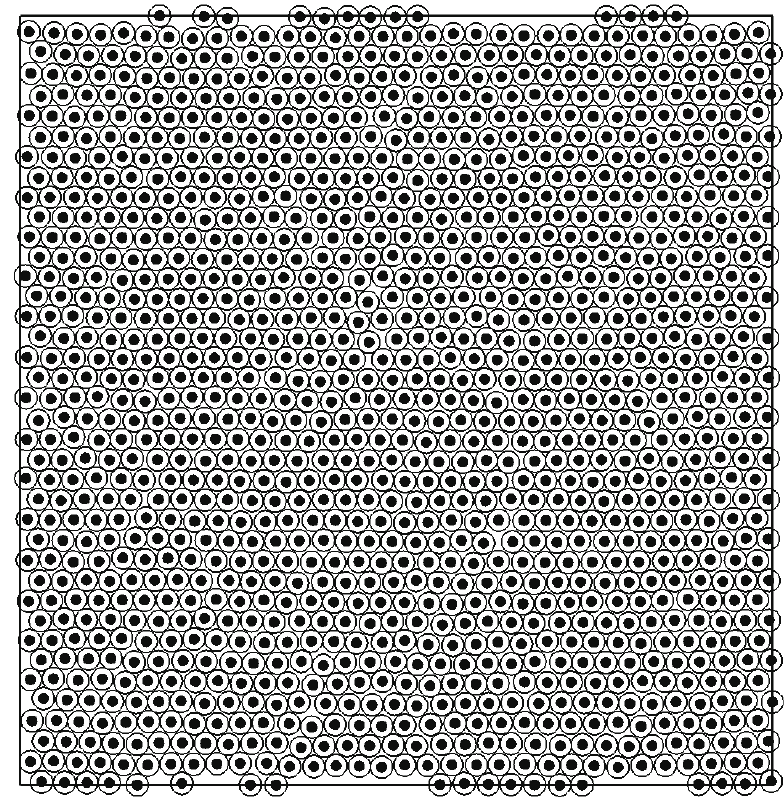}
	\caption{Snapshot of the model polymer-grafted colloidal system in $\mu PT$-ensemble at $( k_B T / \epsilon_0 = 0.1, P \sigma_1^{2} / \epsilon_0 = 0.45, \mu' = 18.42 )$ and $N_0 = 1089$ at $9 \times 10^6$ MCS. The initial system volume is set at $V / \sigma_1^2 = 4778.38$. Black circles represent the inner cores of the particles and white ones denote the outer cores.
	}
	\label{fig:Frog2DmuPTS20RD01e09P045T01IniN1089IniV4778.38_009000000MCS}
\end{figure}
\begin{figure}[!tb]
	\centering
	\includegraphics[clip]{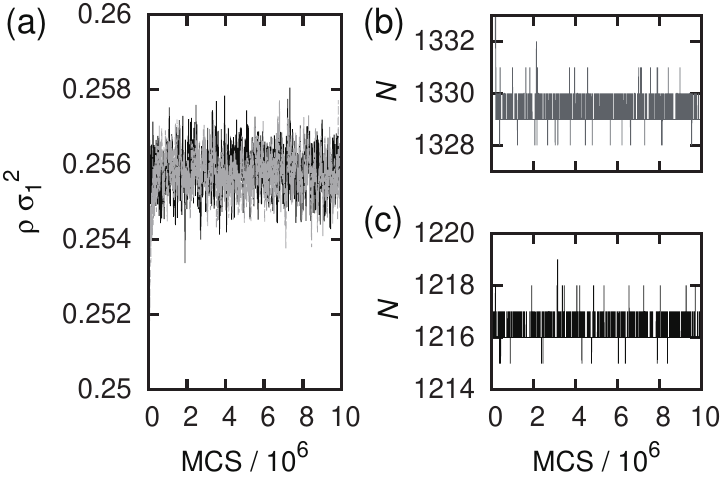}
	\caption{Time evolution of $\rho \sigma_1^2$, (a), and $N$, (b) and (c), in the simulations with the $\mu PT$-ensemble at $( k_B T / \epsilon_0 = 0.1, P \sigma_1^{2} / \epsilon_0 = 0.45, \mu' = 18.42 )$. Black lines denote the results of that are started from the initial condition with $V / \sigma_1^2 = 4778.38$ and $N_0 = 1089$ and grey lines from that with $V / \sigma_1^2 = 5647.18$ and $N_0 = 1221$.}	\label{fig:Frog2DmuPTS20RD01e09P045T01IniN1089IniV4778.38IniN1221IniV5647.18-RhoAndN}
\end{figure}

Simulation with the $\mu PT$-ensemble is started from an initial state with $V / \sigma_1^2 = 4778.38$ and $N_0 = 1089$. We prepared this initial configuration by removing the particles from the equilibrium configuration with $33 \times 38 = 1254$ particles arranged on a homogeneous triangular lattice, which results in an inhomogeneous particle configuration.\footnote[2]{We remove the particles from the columns beside the right edge of the system box until the number of the particles becomes $N_0$, where the $33 \times 38 = 1254$ particles are arranged similar to Fig.~\ref{fig:Frog2DmuPTS20RD01e09P045T01IniN1089IniV4778.38_009000000MCS}.} Such an inhomogeneous configuration is swiftly equilibrated in $\mu PT$-ensemble as is shown in Fig.~\ref{fig:Frog2DmuPTS20RD01e09P045T01IniN1089IniV4778.38_009000000MCS}. A snapshot of the system simulated with the $\mu PT$-ensemble, which is presented in Fig.~\ref{fig:Frog2DmuPTS20RD01e09P045T01IniN1089IniV4778.38_009000000MCS}, shows the defect-free triangular crystal. Temporary small defects due to thermal fluctuation are sometimes found in the system, whereas long-lived defects are absent. Time evolution of $\rho \sigma_1^2$ and $N$ is plotted in Fig.~\ref{fig:Frog2DmuPTS20RD01e09P045T01IniN1089IniV4778.38IniN1221IniV5647.18-RhoAndN}. Small fluctuation in $N$ indicates the high stability of this defect-free crystalline state. Different initial particle configuration also results in a similar defect-free crystalline state with the same average $\rho \sigma_1^2$. Time evolutions of $\rho \sigma_1^2$ and $N$ for simulation on the system with different $N_0$ and initial $V / \sigma_1^2$ are also plotted in Fig.~\ref{fig:Frog2DmuPTS20RD01e09P045T01IniN1089IniV4778.38IniN1221IniV5647.18-RhoAndN}, which show the same defect-free structure. This demonstrates the high stability of the defect-free crystalline state obtained via three-reservoirs method.

Next, we compare physical characteristics of $\mu PT$-ensemble with these of the conventional ensembles. For this purpose, we perform simulations also with the conventional ensembles, \textit{i.e.} $NPT$ and $\mu VT$-ensembles, with the same parameters $( k_B T / \epsilon_0 = 0.1, P \sigma_1^{2} / \epsilon_0 = 0.45, \mu' = 18.42 )$.

\begin{figure}[!tb]
	\centering
	\includegraphics[clip]{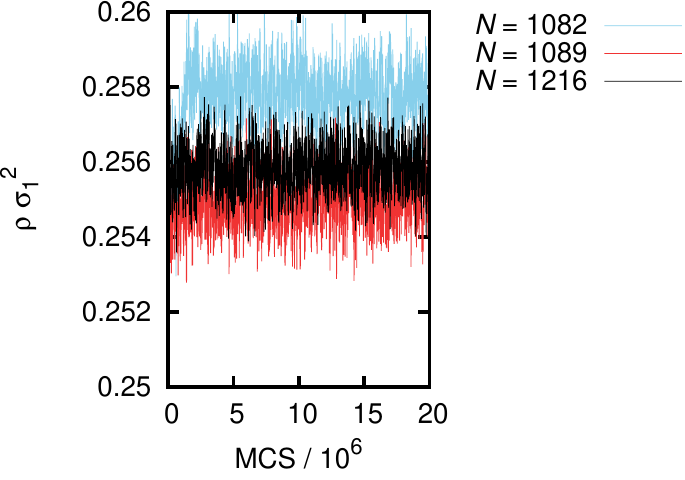}
	\caption{Time evolution of $\rho \sigma_1^2$ in $NPT$-ensemble at $( k_B T / \epsilon_0 = 0.1, P \sigma_1^{2} / \epsilon_0 = 0.45 )$.}
	\label{fig:Frog2DNPTS20P045T01N-Rho}
\end{figure}

\begin{figure*}[!tb]
	\centering
	\includegraphics[clip]{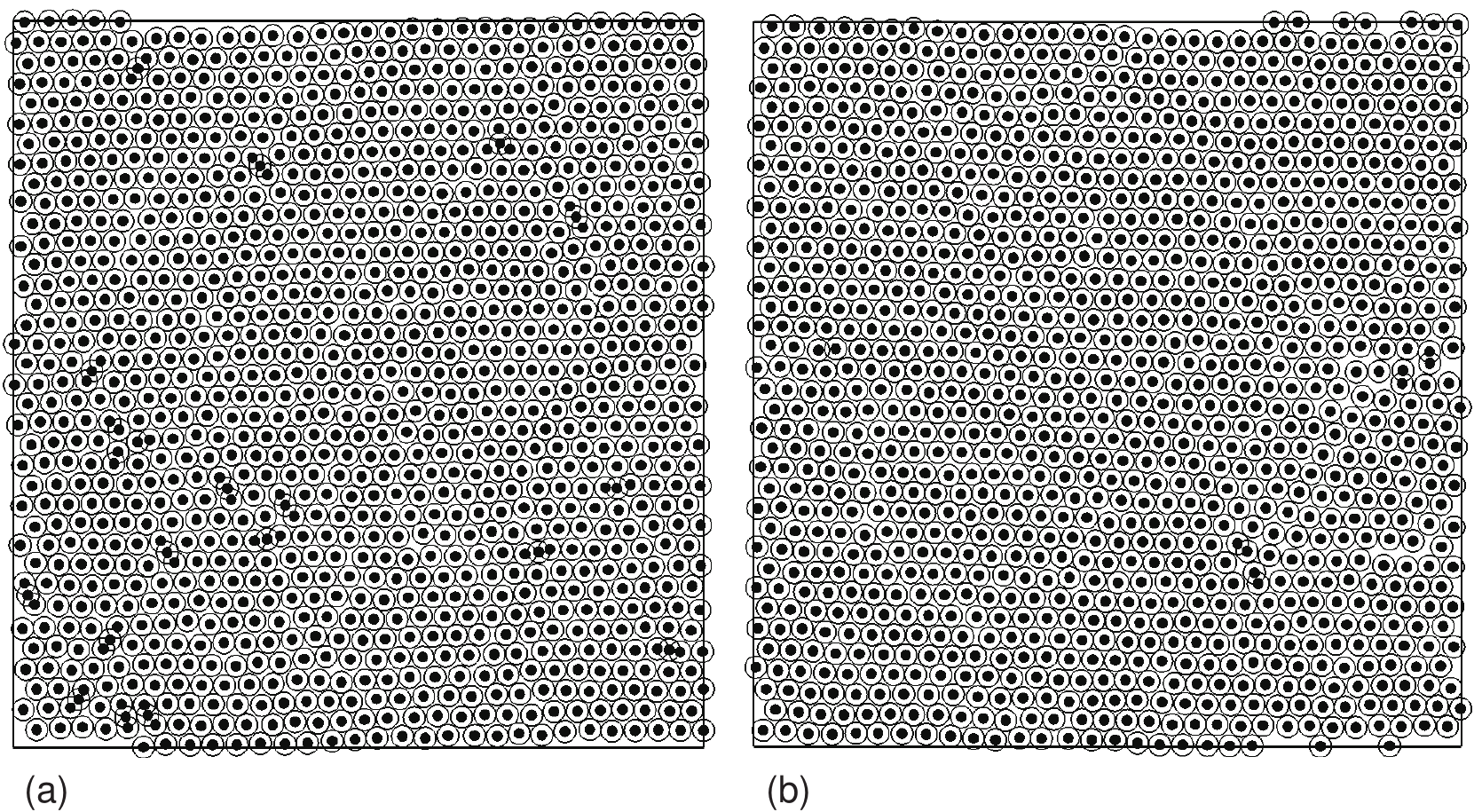}
	\caption{Snapshots of the model polymer-grafted colloidal system simulated with $NPT$-ensemble at $( k_B T / \epsilon_0 = 0.1, P \sigma_1^{2} / \epsilon_0 = 0.45 )$ and $2 \times 10^7$ MCS. Black circles represent the inner cores of the particles and white ones denote the outer cores. (a): $N = 1082$. (b): $N = 1089$.}
	\label{fig:Frog2DNPTS20P045T01N1082N1089_020000000MCS}
\end{figure*}

Different from the $\mu PT$-ensemble case, in the present $NPT$-ensemble case, we have to manually tune the value of $N$ so that the perfect ordered equilibrium structure can be obtained. For this reason, we perform a series of simulations with $NPT$-ensemble for all the values of $N$ within an interval $1070 \le N \le 1089$, where $p_V = 1 / N$ and 1 MCS = $N$ simulation steps are fixed and the initial system volume is still $V / \sigma_1^2 = 4778.38$. The only parameter that is changed from the above $\mu PT$-ensemble simulation is $N$. Typical examples of the time evolution of $\rho \sigma_1^2$ in these simulation runs are plotted in Fig.~\ref{fig:Frog2DNPTS20P045T01N-Rho}. For reference, the result with the parameter optimized in $\mu PT$-ensemble, $N = 1216$, is also plotted in this figure. The system with this optimized parameter shows the defect-free triangular crystal and its value of $\rho \sigma_1^2$ is close to the results of the $\mu PT$-ensemble. However, with any value of $N$ in the above interval, long-lived defects, mostly point defects, appear in the system, as is shown in Fig.~\ref{fig:Frog2DNPTS20P045T01N1082N1089_020000000MCS}. Moreover, the average value of $\rho \sigma_1^2$ changes with the change in $N$ slightly, nonmonotonically, and sensitively. This behavior is different from the results of the $\mu PT$-ensemble. This means that, although the external intensive variables $(T, P)$ are fixed, physical properties of the system in $NPT$-ensemble sensitively depend on the external extensive variable, $N$.

\begin{figure}[!tb]
	\centering
	\includegraphics[clip]{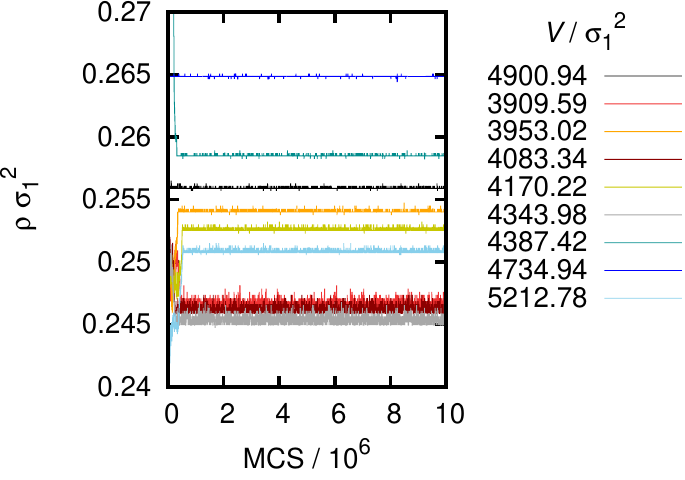}
	\caption{Time evolution of $\rho \sigma_1^2$ obtained with $\mu VT$-ensemble at $( k_B T / \epsilon_0 = 0.1, \mu' = 18.42 )$. The black line denotes the result at $V / \sigma_1^2 = 4900.94$, which is the optimized value obtained with three-reservoirs method.}
	\label{fig:Frog2DmuVTS20RD01e09T01IniV-Rho}
\end{figure}

In $\mu VT$-ensemble, the system size has to manually be tuned. Actually, we try such a manual tuning by performing the $\mu VT$-ensemble simulation with various system sizes in an interval $3909.59 \le V / \sigma_1^2 \le 5212.78$. Aspect ratio of the system box is kept at a typical value for molecular simulation, $L_x / L_y = 1.00$. Simulation parameters that are changed from those of the corresponding $\mu PT$-ensemble simulation are $V$ and $N_0 = 33 \times 38 = 1254$. We assume that 1 MCS = $N_0$ simulation steps. All the systems we have simulated show the defect-free triangular crystals, since point defects are directly removed by the particle insertion and deletion processes. Long-lived defects are not found in these simulation runs with $\mu VT$-ensemble. However, the average value of $\rho \sigma_1^2$ sensitively and nonmonotonically depends on the value of $V / \sigma_1^2$. Typical examples of time evolution of $\rho \sigma_1^2$ are given in Fig.~\ref{fig:Frog2DmuVTS20RD01e09T01IniV-Rho}. Although the intensive variables $(T, \mu)$ are specified by the reservoirs, physical properties of the system in $\mu VT$-ensemble significantly depend on the extensive variable $V$. The data for the simulation with $V / \sigma_1^2 = 4900.94$, which is the optimized value obtained in the three-reservoirs method, are also plotted by a black line. We can confirm that the optimized value of $\rho \sigma_1^2$ is preserved during the simulation run. This result justifies the simulation results obtained using three-reservoirs method.

In the $\mu VT$-ensemble simulations, the particle density $\rho$ can be adjusted only discretely because of the discrete nature of $N$, and therefore the behavior of $\rho$ in $\mu VT$-ensemble is rather abrupt and sensitive to the other parameters. However, in this $\mu VT$-ensemble case, the point defects can rather easily be removed even in high-density states \textit{e.g.} in a triangular crystal. On the other hand in $NPT$-ensemble, $\rho$ is changed by the change in the continuous dynamic variables, $(L_x, L_y, L_z)$. This results in a smaller change in the average $\rho$ compared with that in $\mu VT$-ensemble. However, long-lived defects are frequently found in $NPT$-ensemble. Three-reservoirs method overcomes these disadvantages of $\mu VT$ and $NPT$-ensembles and, at the same time, it inherits the advantages of these ensemble methods. In $\mu PT$-ensemble, both point defects and line defects can be removed easily, and the extensive variables can finely and spontaneously be tuned.

\subsection{Global equilibrium}
\label{subsec:GlobalEquilibrium}
The simulation results in section~\ref{subsec:ExaminationOf3-reservoirsMethod} show that it is a tedious task to perform the manual tuning of $N$ and the system size in conventional ensembles, whereas these extensive variables are automatically and finely tuned in our $\mu PT$-ensemble. In the conventional ensembles where at least one extensive variable is fixed, physical properties of the system are significantly dependent on the value of such a fixed extensive variable. This illustrates that, for example, the equilibrium state of a system in $NPT$-ensemble depends on $N$ even though the intensive variables $T$ and $P$ are fixed. As another example, the equilibrium state of a system in $NVT$-ensemble changes with $N$ and $(L_x, L_y, L_z)$ even though the external intensive variables $T$ and $\rho = N/V$ are fixed. The equilibrium state in conventional ensembles is specified by the parameter sets of the external extensive variables as well as the external intensive variables. In the present article, we tentatively define the local equilibrium state as the equilibrium state at each parameter set of the external extensive variables with the fixed external intensive variables.

In $\mu PT$-ensemble, however, the extensive variables are finely and spontaneously tuned and the most stable state over the local equilibrium states at the given $(T, P, \mu)$ is automatically obtained. In the present article, we tentatively call this equilibrium state in $\mu PT$-ensemble the global equilibrium state.

As these dependences of the physical properties on the external extensive variables in conventional ensembles are regarded as a finite size effect, the three-reservoirs method is a technique to remove the finite size effect of the conventional ensemble methods. Any local equilibrium states are, in the thermodynamic limit, identical to the global equilibrium state.

\subsection{Entropy and free energy calculation in $\mu PT$-ensemble}
\label{subsec:EntropyAndFreeEnergyCalculationInMuPTensemble}
In molecular simulations, physical quantities of the simulation system are defined through the ensemble average over the probability distribution of the microstates in the phase space. The evaluation of the free energy of such a system, however, is equivalent to the evaluation of its partition function. As the partition function is not an averaged quantity over the ensemble, its evaluation demands a high dimensional integral over the whole phase space, resulting in unrealistically large amounts of computational cost.

Instead of evaluating the partition function directly, a derivative of the free energy with a control parameter, \textit{e.g.} pressure in the canonical ensemble, is calculated for the sake of evaluating the free energy in standard molecular simulation~\cite{Frenkel:UnderstandingMolecularSimulation2002}. This technique, which is usually called the thermodynamic integration, gives the free energy difference between two different thermodynamic states.

However, with this technique one cannot go across a first-order phase transition. For example, when the two thermodynamic states are located in a solid phase and in a fluid phase respectively, a first-order transition occurs in the middle of the integration path. Due to possible hysteresis at this transition point, forward and backward integration paths between the two states in general gives different values for the free energy difference. This problem also affects the other free energy calculation techniques, \textit{e.g.} histogram reweighting technique~\cite{Ferrenberg:1989,Kumar:1992} and the method of expanded ensembles~\cite{Lyubartsev:1992}. In order to overcome this difficulty, we need to find appropriate reference states, whose free energy values are already known~\cite{Frenkel:UnderstandingMolecularSimulation2002,Ueda:MolecularSimulationFromClassicalToQuantumMethods}, \textit{e.g.} Einstein solid for the free energy calculation of crystals~\cite{Frenkel:1984,DoctoralThesis}. These reference states provide reversible integration paths. If we cannot find such reference states, we have to find integration paths that bypasses the first-order transition line, \textit{e.g.} an integration path that is arranged with the help of artificially introduced external fields~\cite{Sheu:1995,Mueller:2009PCCP,Norizoe:2010Faraday,DoctoralThesis}. After setting a reversible integration path using these techniques, we run simulations at a large number of state points along the integration path, and integrate the derivative of the free energy along this path. In addition to the discretization error in the integration along the path, occurrence of defects in the system also affects the accuracy of the free energy evaluation of the ordered structures.

Here we propose, based on Euler equation in thermodynamics, a convenient method for the free energy calculation using the systems connected to the three reservoirs. The entropy of the system per particle, denoted by $s = S / N$, satisfies Euler equation,
\begin{equation}
\label{eq:EulerEquationPerParticle}
	s( T, P, \mu ) = \left\langle \frac{ 1 }{ TN } \left( PV - \mu N + U + E_{\text{k}} \right) \right\rangle _{T, P, \mu},
\end{equation}
where $E_{\text{k}}$ denotes the total kinetic energy of the system, $U$ denotes the total potential energy, and $\langle \dotsm \rangle _{T, P, \mu}$ is the ensemble average at $( T, P, \mu )$. In the right-hand side of this equation, the 3 intensive variables, $T$, $P$, and $\mu$ of the system, relax to the equilibrium values that are equal to those given by the reservoirs. $\langle E_{\text{k}} / N \rangle _{T, P, \mu}$ is determined via the equipartition theorem, \textit{e.g.} $(3/2) k_B T$ for monatomic molecules. The ensemble averages $\langle V / N \rangle _{T, P, \mu}$ and $\langle U / N \rangle _{T, P, \mu}$ can directly and readily be evaluated through the simulation runs of the three-reservoirs method. Therefore, according to this Euler equation, $s( T, P, \mu )$ is determined from our simulation at one state point $( T, P, \mu )$, which means that this free energy evaluation method requires far smaller amounts of computation than the other standard methods do. The free energy of the system, \textit{e.g.} Helmholtz free energy, Gibbs free energy, and grand potential, is also obtained from this result in a similar manner. Since our entropy and free energy calculation method is free from any thermodynamic integration paths, the first-order phase transition does not affect our evaluation method. In addition, with the three-reservoirs method we can easily eliminate the defects which is the main origin of the error in the free energy evaluation for the ordered phases. This raises the accuracy of the evaluation.

In our molecular simulation, $\mu$ includes the thermal de Broglie wave length $\varLambda$, as is discussed in appendix~\ref{sec:MolecularMCTechniqueInMuVTAndNPT-ensembles}, \textit{i.e.} eq.~\eqref{eq:ChemicalPotentialOfRealParticles}. $h$ and $m$ in $\varLambda$ cancel when the free energy difference between two different state points is calculated. Therefore, we do not have to take care of these $h$ and $m$.

Although the above method based on Euler equation is, in principle, applicable to simulations with the other ensembles, \textit{e.g.} $NVT$ and $NPT$-ensembles, one has to measure intensive variables $P$ and/or $\mu$ as ensemble averaged values. Such a procedure requires a computationally expensive analysis. For example, the measurement of $\mu$, \textit{i.e.} Gibbs free energy per particle, demands a vast amount of simulation, especially in high particle density regions~\cite{Widom:1963,Shing:1982cpd,Rosenbluth:1955,Frenkel:1992,Pablo:1992,Norizoe:2010Faraday,DoctoralThesis}. However, in the $\mu VT$-ensemble or $\mu PT$-ensemble, we do not have to evaluate $\mu$ because it is already given by the reservoir. Same is true for the evaluation of the pressure $P$. As a result, with the use of the $\mu PT$-ensemble, we can skip tedious evaluations of the intensive variables because all the essential intensive variables $\mu$, $P$, and $T$ are already specified by the reservoirs.

In addition, metastable structures and defects, which frequently appear in conventional ensembles, affect the results of this entropy and free energy evaluation. When the system is in metastable states or outside the global equilibrium state, the intensive variables given from the reservoirs are inconsistent with the values of these variables in the simulation system. This means that both $\mu$ and $P$ need to be analyzed in the simulation rather than to use the specified value by the reservoirs. However, as the defects are quickly eliminated and the system reaches the global equilibrium in the simulations with the $\mu PT$-ensemble, the evaluation of the free energy is free from the above problem associated with the metastable states and the local equilibrium states.

\subsubsection{Construction of equilibrium phase diagrams.~~}
\label{subsubsec:ConstructionOfEquilibriumPhaseEiagrams}
When one tries to construct the equilibrium phase diagrams using conventional simulation methods, candidates for the equilibrium structure have to be chosen prior to the simulation and the free energy of each candidate should be measured and compared with each other with high accuracy, for example, with a typical error level~\cite{Norizoe:2010Faraday,DoctoralThesis,Fomin:2008} of $10^{-4}$ to $10^{-6}$. A variety of phases, \textit{e.g.} crystals, the string-like assemblies, and the other ordered and disordered structures should be considered the candidates for the equilibrium phase. In actual calculation, we empirically select some of these potential candidates and discard the others. However, we cannot deny the possibility that we have discarded the true equilibrium structure in this selection process. In addition, there is another possibility that the equilibrium structure is a totally new structure which has not been discovered yet.\footnote[3]{Furthermore, in conventional ensembles, physical properties of the system are significantly and sensitively dependent on the external extensive variables even though the external intensive variables are fixed. Even if the free energy is precisely measured in conventional ensembles for the correct candidates, obtained results are restricted to the local equilibrium at the given parameter set of the external extensive variables, \textit{i.e.} a small region of the phase space including the external extensive variables. The free energy densities measured in one local equilibrium state change in another local equilibrium state at the same external intensive variables. This could result in distinct structures in each local equilibrium state. These results illustrate a challenge to find and confirm the global equilibrium state, \textit{i.e.} the most stable state over the local equilibrium states, in simulation of conventional ensembles. For example, even a disordered fluid could appear in the global equilibrium, while a crystal in the local equilibrium.}

With the use of $\mu PT$-ensemble, however, one can obtain the global equilibrium structure directly as a result of the fine tuning of the extensive variables. Therefore, the above-mentioned problem encountered in the construction of the phase diagram using the standard ensembles can be avoided when we use $\mu PT$-ensemble.

\section{Conclusions}
\label{sec:Conclusions}
We have studied thermodynamics, statistical mechanics, and molecular MC simulation algorithms of $\mu PT$-ensemble. Guggenheim formally introduced Boltzmann factor (statistical weight) of this ensemble with an assumption that the averages of extensive variables can be determined~\cite{Guggenheim:1939}. However, this assumption contradicts the indetermination of extensive variables in $\mu PT$-ensemble. In addition, other characteristics of this ensemble were totally absent in Guggenheim's discussion and other early works~\cite{Prigogine:1950,Hill:StatisticalMechanicsPrinciplesAndSelectedApplications,Sack:1959}. These early researchers concentrated on the formalism, i.e. mathematical aspects of the partition function of this ensemble, only at the thermodynamically stable point~\cite{Koper:1996}. Physical aspects of the ensemble have not seriously been discussed. In the present work, we have shed light on these problems and have discovered thermodynamic and statistical mechanical characteristics of $\mu PT$-ensemble, \textit{e.g.} the thermodynamically stable point, thermodynamic degrees of freedom, thermodynamics outside the thermodynamically stable point, maximization of entropy density, quick equilibration due to shortcuts in the phase space, \textit{etc.} We have also shown that $\mu PT$-ensemble is built as a combination of the 3 ensembles each of which is combined to 2 reservoirs, \textit{i.e.} $NPT$, $\mu VT$, and $\mu PS$-ensembles. The 3 corresponding free energy densities, \textit{i.e.} Gibbs free energy per particle, grand potential per volume, and $(E - \mu N + PV) / S$, are simultaneously minimized in $\mu PT$-ensemble, rather than the thermodynamic potential of $\mu PT$-ensemble, denoted by $\Phi$.

We have proposed a molecular MC simulation method based on $\mu PT$-ensemble (three-reservoirs method)~\cite{2010:NorizoeMuPTLetterArXiv}. We can show that this three-reservoirs method gives a physically acceptable ensemble, which allows us to trace the physical trajectories in the phase space. Since three-reservoirs method is built as a combination of conventional $NPT$ and $\mu VT$-ensembles, programming is lighter than other advanced techniques. In addition, the thermodynamically stable point is determined according to Gibbs-Duhem equation in a short computational time. These features mean that three-reservoirs method requires a small amount of preparation and that we can quickly start production simulation runs, although other advanced techniques demand a large quantity of complicated preparation, \textit{e.g.} advanced programming, the precise adjustments of the artificial weights necessary for multicanonical technique, and accurate free energy measurement essential for the expanded ensemble technique~\cite{Lyubartsev:1992}.

These advantages over other simulation techniques facilitate and reduce the total work flow of our three-reservoirs method compared with the conventional methods such as $NPT$, $\mu VT$, or multicanonical method. Furthermore, only with the three-reservoirs method, we can \textit{simultaneously} and \textit{automatically} tune the number of particles $N$ and the system size to obtain the equilibrium ordered state. This unique advantage of the three-reservoirs method could enhance the understanding of those systems that were obtained via the other standard simulation techniques. For example, for perfect crystals, both $N$ and $\left( L_x, L_y, L_z \right)$ must be integer multiples of the unit structure of the crystal structure, which is in general not known \textit{a priori}. In principle, by measuring the free energy density of various structures at each set of $N$ and $\left( L_x, L_y, L_z \right)$, these extensive parameters can manually be tuned in simulation of $NPT$-ensemble, $\mu VT$-ensemble, multicanonical ensemble, \textit{etc.} In practice, however, this manual tuning requires much computational effort. On the other hand, with the use of our three-reservoirs method, we can automatically achieve such optimization.

For a solid at finite temperature $T$, the system with the $\mu PT$-ensemble reaches a globally-anisotropic defect-free ordered state as the equilibrium state by crossing the metastable states through the shortcuts in the phase space due to the additional degrees of freedom. On the other hand, in conventional ensembles, physical properties of the system sensitively and discretely depend on $N$ and/or $\left( L_x, L_y, L_z \right)$ even though external intensive variables are fixed. This results in a requirement of the manual tuning of these extensive variables to obtain the global equilibrium state.

These results illustrate that our method can be applied to a variety of physical systems for the sake of studying ordered structures in equilibrium at finite $T$, \textit{e.g.} lamellae composed of diblock copolymers, smectic phase of liquid crystals, and fluid bilayer membranes. Three-reservoirs method can also be applied to numerical calculation of the equation of state that relates $\mu$, $P$, and $T$. We have also shown that the entropy and the free energy can quickly be evaluated in $\mu PT$-ensemble based on Euler equation. This feature is essentially important in the construction of the phase diagram of condensed materials.

\section*{Acknowledgments}
The authors wish to thank Professor Komajiro Niizeki and Mr Masatoshi Toda for helpful suggestions and discussions.
We also thank the anonymous referees for their valuable suggestions and comments.
This work is partially supported by a grant-in-aid for science from the Ministry of Education, Culture, Sports, Science, and Technology, Japan.

\appendix
\section{Molecular MC technique in $\mu VT$ and $NPT$-ensembles}
\label{sec:MolecularMCTechniqueInMuVTAndNPT-ensembles}
In this appendix, we show molecular MC simulation technique~\cite{Ueda:MolecularSimulationFromClassicalToQuantumMethods,Cates:SoftAndFragileMatter,Frenkel:UnderstandingMolecularSimulation2002,ComputerSimulationOfLiquids} for single component systems based on the $\mu VT$ and $NPT$-ensembles. We use the following notations: $\rVector_i$ denotes the spatial coordinates of the particle $i$, $m$ mass of a particle, $k_B T$ the thermal energy, $U$ the potential energy of the system, and $h$ Planck's constant. The chemical potential of a system consisting of real particles in the canonical ensemble ($NVT$-ensemble) is given by,
\begin{equation}
\label{eq:ChemicalPotentialOfRealParticles}
	\frac{ \mu_{\text{real}} (T, \rho) }{ k_B T }
		:= \frac{ \mu_{\text{ideal}} (T, \rho) }{ k_B T } + \frac{ \mu_{\text{excess}} (T, \rho) }{ k_B T }, \quad \rho := N / V,
\end{equation}
where
\begin{equation}
	\label{eq:ChemicalPotentialOfIdealParticles}
	\frac{ \mu_{\text{ideal}} (T, \rho) }{ k_B T } := \log \left( \varLambda^3 \rho \right),
\end{equation}
and
\begin{equation}
	\label{eq:ThermalDeBroglieWaveLength}
	\varLambda := \frac{h}{ \sqrt{2 \pi m k_B T} }.
\end{equation}
Here, $\mu_{\text{ideal}} (T, \rho)$ denotes the chemical potential of an ideal gas in $NVT$-ensemble and $\varLambda$ is called the thermal de Broglie wave length. The quantity $\mu_{\text{excess}} (T, \rho)$ denotes the excess chemical potential that is originated from the interaction between the real particles~\cite{Frenkel:UnderstandingMolecularSimulation2002,DoctoralThesis}.

Simulation results in $\mu VT$-ensemble are independent of $\varLambda$. This means that $\varLambda$ only appears in the chemical potential $\mu$ at the reference point in $\mu VT$-ensemble simulation, which is discussed in section~\ref{subsec:MCSimulationMethodInMuVTensemble}.

\subsection{MC simulation method in $\mu VT$-ensemble}
\label{subsec:MCSimulationMethodInMuVTensemble}
In the present section, we discuss the MC method of a single component system in the grand canonical ensemble, which is also called $\mu VT$-ensemble because $\mu$, $V$, and $T$ are kept fixed.

In simulation with the $\mu VT$-ensemble~\cite{Norman:1969,Ueda:MolecularSimulationFromClassicalToQuantumMethods,Frenkel:UnderstandingMolecularSimulation2002}, particles are inserted into and deleted from the system in addition to Metropolis trial displacement of particles. In one simulation for this ensemble, these steps are included in:
\begin{enumerate}[i)]
	\item with probability $p_G / 2$, trial particle insertion into the system
	\item with probability $p_G / 2$, trial particle deletion from the system
	\item with probability $1 - p_G$, trial displacement based on Metropolis algorithm, \textit{i.e.} perturbation to one particle
\end{enumerate}
is chosen, where $p_G$ is a constant fixed in an interval $0 \le p_G \le 1$. Algorithms of trial particle insertion and deletion are discussed in the following.

\subsubsection{Particle insertion.~~}
\label{subsubsec:ParticleInsertion}
We assume that one particle is inserted into the system that is composed of $N$ particles. The position of this inserted particle $\rVector_{N+1}$ is chosen uniformly over the system box. The coordinates of the $N$ particles, $( \rVector_1 , \dots , \rVector_N )$, are fixed during this particle insertion step. The state after the insertion, \textit{i.e.} the state of $N+1$ particles, is accepted as a new state with the probability,
\begin{align}
\label{eq:AcceptanceCriterionParticleInsertion}
	& \text{acc} ( N \to N+1 )  \notag \\
	& \qquad = \min \left( 1, \frac{ V }{ \varLambda^3 (N + 1) } \exp \left[ - \frac{1}{ k_B T } U^{\text{ins}}_{\text{excess}} + \frac{\mu}{k_B T} \right] \right),  \\
	& U^{\text{ins}}_{\text{excess}} := U \left( \rVector_1 , \dots , \rVector_{N+1} \right) - U \left( \rVector_1 , \dots , \rVector_N \right).  \notag
\end{align}
A function $\min( a_1, a_2 )$ returns the smaller of two arguments, $a_1$ and $a_2$. If the trial insertion is rejected, the state before the insertion is kept for the next simulation step.

\subsubsection{Particle deletion.~~}
\label{subsubsec:ParticleDeletion}
We assume that one particle is randomly chosen and attempts to be removed from the present system composed of $N$ particles. This chosen particle, denoted by index $j$, is removed from the system with the probability,
\begin{gather}
\label{eq:AcceptanceCriterionParticleDeletion}
	\text{acc} ( N \to N-1 ) = \min \left( 1, \frac{ \varLambda^3 N }{ V } \frac{ 1 }{ \exp \left[ - \frac{ 1 }{ k_B T } U^{\text{del}}_{\text{excess}} + \frac{\mu}{k_B T} \right] } \right),  \\
	U^{\text{del}}_{\text{excess}} := U \left( \rVector_1 , \dots , \rVector_N \right) - U \left( \rVector_1 , \dots , \rVector_{j-1}, \rVector_{j+1}, \dots , \rVector_N \right).  \notag
\end{gather}
If the trial deletion is rejected, the state before the deletion is kept for the next simulation step.

When $\mu_{\text{real}}$ is substituted for $\mu$ in eqs.~\eqref{eq:AcceptanceCriterionParticleInsertion} and \eqref{eq:AcceptanceCriterionParticleDeletion}, these acceptance criteria are independent of $\varLambda$. This illustrates that simulation results in $\mu VT$-ensemble are free from the actual value of $\varLambda$. Therefore, $\varLambda$ appears only in the chemical potential $\mu$ at the reference point in the $\mu VT$-ensemble simulation~\cite{DoctoralThesis}.

\subsection{MC simulation method in $NPT$-ensemble}
\label{subsec:MCSimulationMethodInNPTensemble}
MC simulation method in isobaric-isothermal ensemble~\cite{Wood:1968,Wood:1970,McDonald:1972,Frenkel:UnderstandingMolecularSimulation2002}, also called $NPT$-ensemble, is briefly summarized in this section. In addition to $N$ and $T$, the system pressure, denoted by $P$, is given from the outside, in this ensemble.

In simulation of $NPT$-ensemble, the system size is changed, in addition to the trial move of particles. In one simulation step, we include the following steps:
\begin{enumerate}[i)]
	\item with probability $p_V$, trial system size change
	\item with probability $1 - p_V$, trial displacement based on Metropolis algorithm, \textit{i.e.} perturbation of one particle
\end{enumerate}
is chosen, where $p_V$ is a constant fixed in an interval $0 \le p_V \le 1$. Algorithms of trial system size change are discussed in the following.

\subsubsection{Trial system size change in $NPT$-ensemble.~~}
\label{subsubsec:TrialSystemSizeChangeInNPTensemble}
To change the system box size, we use the following 4-step algorithm.

We assume that the system box size, denoted by $\left( L_x, L_y, L_z \right)$, is changed to a new system size, $\left( L'_x, L'_y, L'_z \right)$. Unlike the conventional MC simulations in $NPT$-ensemble based on McDonald's method~\cite{Frenkel:UnderstandingMolecularSimulation2002}, each element of the system size $\left( L_x, L_y, L_z \right)$ is independently changed in our algorithm.
\begin{enumerate}[1)]
	\item The new system size is chosen,
	\begin{gather}
	\label{eq:NewSystemSizeInNPTensemble}
		L'_x = L_x + \varDelta L (1 - 2 \xi_x),  \notag \\
		L'_y = L_y + \varDelta L (1 - 2 \xi_y),			\\
		L'_z = L_z + \varDelta L (1 - 2 \xi_z),  \notag
	\end{gather}
	where $\varDelta L$ is a small constant length, $\xi_x, \xi_y$, and $\xi_z$ are random numbers uniformly distributed over an interval $0 \le \xi_x, \xi_y, \xi_z \le 1$, and $V' = L'_x L'_y L'_z$ denotes the volume of the new system box.

	\item Coordinates of all the particles before the system size change, $\rVector_i = \left( x_i, y_i, z_i \right)$, are homogeneously scaled,
	\begin{equation}
	\label{eq:NewParticleCoordinatesNPT}
		\rVector'_i = \left(
			( L'_x / L_x ) x_i,
			( L'_y / L_y ) y_i,
			( L'_z / L_z ) z_i \right),
		\qquad  1 \le i \le N.
	\end{equation}
	This is new particle coordinates after the change.

	\item The potential energies of the systems before and after the system size change, $U \left( \rVector_1, \dots , \rVector_N \right)$ and $U \left( \rVector'_1, \dots , \rVector'_N \right)$ respectively, are calculated.

	\item The new system size and particle coordinates are accepted with the probability,
	\begin{multline}
	\label{eq:AcceptanceCriterionVolumeChange}
		\text{acc} ( V \to V' ) =  \\
		\min \left( 1, \exp \left[ - \frac{1}{ k_B T } \Biggl\{ U \left( \rVector'_1, \dots , \rVector'_N \right) - U \left( \rVector_1, \dots , \rVector_N \right) \right. \right. \\
		\left. \left. + P \left( V' - V \right) - N k_B T \log \frac{V'}{V} \Biggr\} \right] \right).
	\end{multline}
	If this trial system size change is rejected, the state before the trial is kept for the next simulation step.
\end{enumerate}

\section{Ensemble average at each $N$ and $V$ in $\mu PT$-ensemble}
\label{sec:EnsembleAverageAtEachNAndVInmuPTensemble}
The ensemble average at each $N$ in $\mu PT$-ensemble is equivalent to the average in $NPT$-ensemble at the same $N$. We illustrate this in the present section.

We assume that a finitely long simulation run is performed in $\mu PT$-ensemble at a thermodynamically stable point. After the equilibration of the simulation system, a finitely large number of microstates of the system are visited. Unnormalized Boltzmann factor of the system is denoted by $w_i$, where the suffix $i$ represents these microstates. $\delta_{lm}$ is Kronecker delta. $\beta = 1 / k_B T$.

After the equilibration of the system, the ensemble average of an intensive physical quantity, $A$, obtained in the present simulation run is:
\begin{align}
\label{eq:EnsembleAveragemuPTForIntensiveQuantity}
	& \langle A \rangle _{T, P, \mu}  \notag \\
	& \quad \cong \sum_{N=0}^{\infty} \frac{ \sum_i \delta_{N_i, N} A_i w_i }{ \sum_i w_i } = \sum_{N=0}^{\infty} \frac{ \sum_i \delta_{N_i, N} w_i }{ \sum_i w_i }  \frac{ \sum_i \delta_{N_i, N} A_i w_i }{ \sum_i \delta_{N_i, N} w_i }  \notag \\
	& \quad = \sum_{N=0}^{\infty} f(N) \frac{ \sum_i \delta_{N_i, N} A_i \frac{1}{ \varLambda^{3N} \, N! } \exp \left[ - \beta \left( P V_i - \mu N + U_i \right) \right] }  { \sum_i \delta_{N_i, N} \frac{1}{ \varLambda^{3N} \, N! } \exp \left[ - \beta \left( P V_i - \mu N + U_i \right) \right] }  \notag \\
	& \quad = \sum_{N=0}^{\infty} f(N) \frac{ \sum_i \delta_{N_i, N} A_i \exp \left[ - \beta \left( P V_i + U_i \right) \right] }  { \sum_i \delta_{N_i, N} \exp \left[ - \beta \left( P V_i + U_i \right) \right] }  \notag \\
	& \quad \cong \sum_{N=0}^{\infty} f(N) \, \langle A \rangle _{T, P, N},
\end{align}
where the summation $\sum_i$ runs over the microstates visited in the simulation, and $f(N)$ is defined as,
\begin{equation}
\label{eq:ProbabilityDensityNInmuPT}
	f(N) := \frac{ \sum_i \delta_{N_i, N} w_i }{ \sum_i w_i },
\end{equation}
and $\langle A \rangle _{T, P, N}$ denotes the ensemble average of $A$ in $NPT$-ensemble. $f(N)$ denotes the occurrence probability of $N$ in the simulation run. An expression similar to eq.~\eqref{eq:EnsembleAveragemuPTForIntensiveQuantity} also holds for the ensemble average of $A$ in $\mu VT$-ensemble. This result illustrates that the short-time average of $A$ in $\mu PT$-ensemble is approximated by the ensemble average in $NPT$-ensemble or $\mu VT$-ensemble.

When the finite size effect of the system, which has been discussed in sections~\ref{subsec:ExaminationOf3-reservoirsMethod} and \ref{subsec:GlobalEquilibrium}, is small, $\langle A \rangle _{T, P, N}$ is dependent on $(T, P)$ and independent of $N$. Therefore, eq.~\eqref{eq:EnsembleAveragemuPTForIntensiveQuantity} yields $\langle A \rangle _{T, P, \mu} = \langle A \rangle _{T, P, N} = A (T, P)$. A similar relation, $\langle A \rangle _{T, P, \mu} = \langle A \rangle _{T, \mu, V} = A (T, \mu)$, also holds, where $\langle A \rangle _{T, \mu, V}$ denotes the ensemble average of $A$ in $\mu VT$-ensemble. Therefore, $\langle A \rangle _{T, P, \mu} = A (T, P) = A (T, \mu)$.

%

\end{document}